\documentclass[fleqn,usenatbib,useAMS]{mnras}
\usepackage{hyperref}
\usepackage{graphicx}	
\usepackage{amsmath}	
\usepackage{amssymb}	
\usepackage{bm}		
\usepackage{pdflscape}	
\usepackage{comment}
\usepackage{overpic,color}
\usepackage{subcaption}
\usepackage{caption}
\usepackage{multicol}        
\usepackage{mathptmx}
\usepackage[T1]{fontenc}
\usepackage{ae,aecompl}
\usepackage{booktabs} 
\usepackage{tablefootnote}
\usepackage{epsfig} 
\def\aj{AJ}                  
\def\apj{ApJ}                  
\def\apjl{ApJL}    
\def\mnras{MNRAS}                  
\def\araa{ARA\& A}
\def\pasp{PASP}
\def\apjs{ApJS}  
\def\aap{A\&A}  
\def\aaps{A\&AS}

\def\pasa{PASA}
\def\nar{New Astron. Rev.}

\defcitealias{Leethochawalit16}{L16}
\newcommand{{\lenstool}}{{\small LENSTOOL}}
\newcommand{\lephare}{{\small LE PHARE}}
\newcommand{\angstrom}{\mbox{\normalfont\AA}}

\title[Monthly Notices of the Royal Astronomical Society]
  {High Resolution spatial analysis of a z $\sim$ 2 lensed galaxy using adaptive coadded source-plane reconstruction}
\author[Sharma et al.]
  {Soniya Sharma$^{1,2}$, Johan Richard$^3$, Tiantian Yuan$^{2,4}$, Anshu Gupta$^{1,2}$, Lisa Kewley$^{1,2}$,
  \newauthor
  Vera Patr\'icio$^5$, Nicha Leethochawalit$^{6}$, Tucker A.Jones$^{7}$ \\
  $^1$Research School of Astronomy and Astrophysics, Australian National University, Cotter Road, ACT 2611, Australia\\
  $^2$ARC Centre of Excellence for All Sky Astrophysics in 3 Dimensions (ASTRO 3D), Australia\\
  $^3$ Univ Lyon, Univ Lyon1, Ens de Lyon, CNRS, Centre de Recherche Astrophysique de Lyon UMR5574, F-69230, Saint-Genis-Laval, France\\
  $^4$ Centre for Astrophysics and Supercomputing, Swinburne University of Technology, Hawthorn, Victoria 3122, Australia \\
 $^5$ Dark Cosmology Centre, Niels Bohr Institute, University of Copenhagen, Juliane Maries Vej 30, DK-2100 Copenhagen, Denmark \\
 $^6$ Cahill Center for Astronomy and Astrophysics, California Institute of Technology, MS 249-17, Pasadena CA 91125, USA \\
 $^7$ Department of Physics, University of California Davis, 1 Shields Avenue, Davis, CA 95616, USA
  }
\date{Released 2018 Xxxxx XX}

\pagerange{\pageref{firstpage}--\pageref{lastpage}} \pubyear{2018}

\def\LaTeX{L\kern-.36em\raise.3ex\hbox{a}\kern-.15em
    T\kern-.1667em\lower.7ex\hbox{E}\kern-.125emX}

\begin{document}
\label{firstpage}

\maketitle
\begin{abstract}
We present spatially resolved analysis of a lensed galaxy, SDSS1958+5950 at $z = 2.225$, from the Cambridge Sloan Survey of Wide Arcs in the Sky (CASSOWARY). We use our new high resolution imaging data to construct a robust lens model for the galaxy group at $z = 0.214$. We employ the updated lens model to combine the Integral Field Spectrographic observations on two highly distorted images of the lensed target. We adopt a forward-modeling approach to deconvolve the effects of point spread function from the combined source-plane reconstruction. The approach is adapted to the lens model magnification and enables a resolution of $\sim$170 pc in the galaxy-source plane. We propose an ongoing merger as the origin of the lensed system on the basis of its source-plane morphology, kinematics and rest-frame emission line ratios. Using our novel technique of adaptive coadded source plane reconstruction, we are able to detect different components in the velocity gradient that were not seen in previous studies of this object, plausibly belonging to different components in the merging system. 
\textit{}
\end{abstract}


\section{Introduction}\label{intro}
Spatially resolved studies of high redshift galaxies, particularly in the peak of galaxy formation epoch $1\lesssim z\lesssim 3$ \citep{Madau14} hold the key for understanding the physics of galaxy formation and evolution. The development of Integral Field Spectrographs (IFS) and Adaptive Optics (AO) Imaging techniques in the last decade have revealed the diverse kinematic state of galaxies in this epoch, ranging from rotationally supported clumpy disks to more dispersion dominated systems \citep[see][for a review]{Glazebrook13}.   

Observationally, AO aided IFS surveys \citep[][]{Forster06, Flores06, Maiolino08, Gnerucci11, Genzel11, Contini12, Forster18} have allowed us to achieve an improved resolution of up to 100 mas ($\sim$ 800 pc at $1\leq z\leq 3$), enabling the sampling of $z \sim$ 1 - 3 galaxies in a few coarse resolution elements. However, in order to understand the physical conditions of H\,{\sc ii }\rm regions and the evolution of ISM properties with redshift, it is imperative to resolve star-forming (SF) regions at high-$z$ in the same manner as we resolve them in the local universe. For example, the physical scales of H\,{\sc ii }\rm regions in the local universe span at least an order of magnitude, from small OB associates ($\sim$ 60 pc), stellar aggregates ($\sim$ 240 pc) to large star complexes ($\sim$ 600 pc), with a decrease in surface brightness from the smallest to the largest scales \citep{Elmegreen06, Gusev14}. Therefore, current high-redshift observations are highly biased towards the H\,{\sc ii }\rm  physics on the largest scales. Similar systematic biases also exist in kinematic and metallicity gradient analysis of high-$z$ galaxies  \citep[e.g.,][] {Yuan13, Jones13, Yuan17}.  

The improved spatial resolution provided by IFS observations of gravitationally lensed galaxies \citep[e.g.][]{Jones10a, Livermore12, Wisnioski15, Livermore15, Leethochawalit16, Mason17, Patricio18} has played a pivotal role in probing the star-forming regions with a physical resolution down to a few hundred parsecs in the galaxy-source plane. Magnification factors of strongly lensed systems can easily range between 1 - 10 \citep[e.g.][]{Richard11}, with reasonable lensing uncertainties \citep[][]{Collett15}. In extreme cases of giant arcs around galaxy groups and galaxy clusters, the magnification can reach up to a few $\times$ 10, rendering a physical resolution of less than 100 pc at $z \sim 2$. Therefore, giant distorted arcs are ideal candidates to study the physics of high-redshift SF regions at highest spatial resolution possible \citep[][]{Swinbank06, Jones10a, Yuan12, Bayliss14, Livermore15, Johnson17, Girard18}. 

While it is the giant arcs that offer the largest magnification, modeling of these arcs is also the most challenging. One of the biggest challenges of studying the giant lensed arcs is the large uncertainty in the lens model. The lensing mass distribution very sensitively controls the accuracy and precision of source-plane reconstructions of these arcs. The current strong lens modeling is unable to match the angular resolution of the imaging data (for e.g. HST $\sim 0.\!\!^{\prime\prime}05$) with a residual RMS of up to a few arcseconds \citep[][]{Limousin07, Lagattuta17, Caminha17}. Line of sight substructures, redshift information of multiple image systems have also been shown to have a significant contribution to the systematics errors arising in strong lens models of lensing clusters \citep{Bayliss14b, Johnson16, Acebron17}.  

However, even with the use of accurate lens models, simple source reconstructions removing the lens deflection from the observations does not allow us to correctly recover the intrinsic source-plane surface brightness. This is because of the point spread function (PSF) convolving with the data in the image plane. PSF remains in the source plane and varies as a function of the source position. The impact of PSF is even more severe in the vicinity of the critical lines. Thus the resulting source-plane resolution limits us from confidently combining independent reconstructions from different multiple images of the same lensed galaxy. In order to fully utilize the power of lensing and recover the physical properties of galaxy delensed, it is important to simultaneously combine the information from different images. This has been shown before by certain case studies \citep[for e.g][]{Coe10, Jones10b, Jones15} of lensed systems.

To easily combine the observations and deal with instrumental PSF, a forward modeling approach is ideal to reconstruct the galaxy on the source plane. The use of forward source modeling techniques has gained increasing popularity in the past decade owing to the fast growing data set of strongly lensed systems. Previous lensing studies \citep{Warren03, Night15, Tess16, Johnson17, Des17} have demonstrated the advantages of these techniques in accurately studying source profiles at scales otherwise unachievable through traditional image inversion methods. However, most of these techniques rely on computationally expensive algorithms to optimize lens model parameters and the extended source simultaneously.

In this work, we conduct a detailed study of $z \sim$ 2 lensed galaxy using a unique forward modeling approach to combine all the available data from different multiple images of the lensed target. The approach is computationally fast and exploits the best-fit lensing mass model in rendering the physical properties from the combined source-plane data. We choose one of the brightest targets in The Cambridge Sloan Survey of Wide Arcs in the Sky (CASSOWARY), SDSS1958+5950 (hereafter referred by its survey-ID: cswa128; $z = 2.225$). CASSOWARY survey presented a large sample of about 100 group-scale gravitationally lensed systems at $z \sim 1 - 3$ from the Sloan Digital Sky Survey \citep[SDSS;][]{York00} imaging data \citep{Stark13}. The survey targeted star-forming galaxies ($z \sim 1 - 3$) being lensed by an early-type galaxy (or group) at $z \sim 0.2 - 0.7$. \cite{Leethochawalit16} (hereafter \citetalias{Leethochawalit16}) presented IFU observations of a representative sample of 15 lensed galaxies from the survey.

cswa128 is lensed by a galaxy group at $z = 0.214$ and reported to have high [N\,{\sc ii}\rm]/H$\alpha$ at the outer edges,  by previous work of (\citetalias{Leethochawalit16}). A lens model based on SDSS imaging data was used to interpret IFU observations on a single lensed arc of this target. In this work, we present high resolution imaging data from the NIRC2 instrument on the Keck I telescope that enables us to build a more complex lens model for this system.

We also discuss new spectroscopic observations on another lensed image of the galaxy than previously observed by \citetalias{Leethochawalit16}. We use the updated lens model to reconstruct two images of cswa128 on the source plane. To overcome the challenges posed by the PSF, we test a forward modeling approach to merge the two IFS data reconstructions on the source plane. Using our novel technique, we find that cswa128 is highly evident of a merging system. 

The paper is organised as follows. In section~\ref{sec:data} we present different observations of the lensed system and an overview of the data reduction process. In section~\ref{sec:lensmodel} we describe the lensing methodology and the new lens model of the galaxy group. In section~\ref{sec:analysis} we detail our technique of coadding different images of the lensed galaxy on the source plane and the resulting source plane properties from the reconstruction. We discuss the overall results and summarize the conclusions under Section~\ref{sec:discussion} of the paper. 

Throughout this paper we adopt a standard $\Lambda$-CDM cosmology with $\Omega_{m} = 0.3$, $\Omega_{\Lambda} = 0.7$ and h $= 0.7$. All magnitudes are given in the AB system \citep{Oke83}. 

\begin{figure*}
\centering
\includegraphics[scale= 0.6,angle=0]{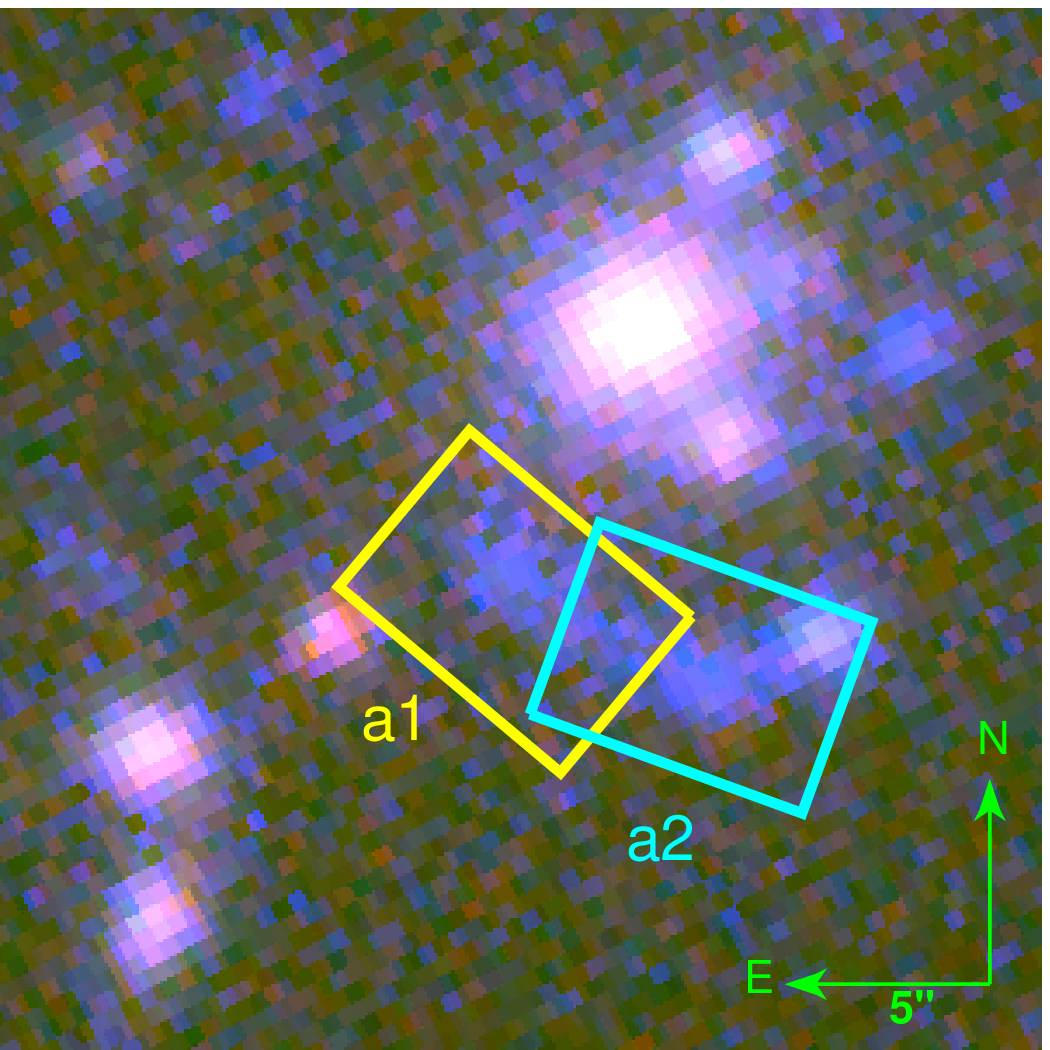}
\includegraphics[scale = 0.6,angle=0]{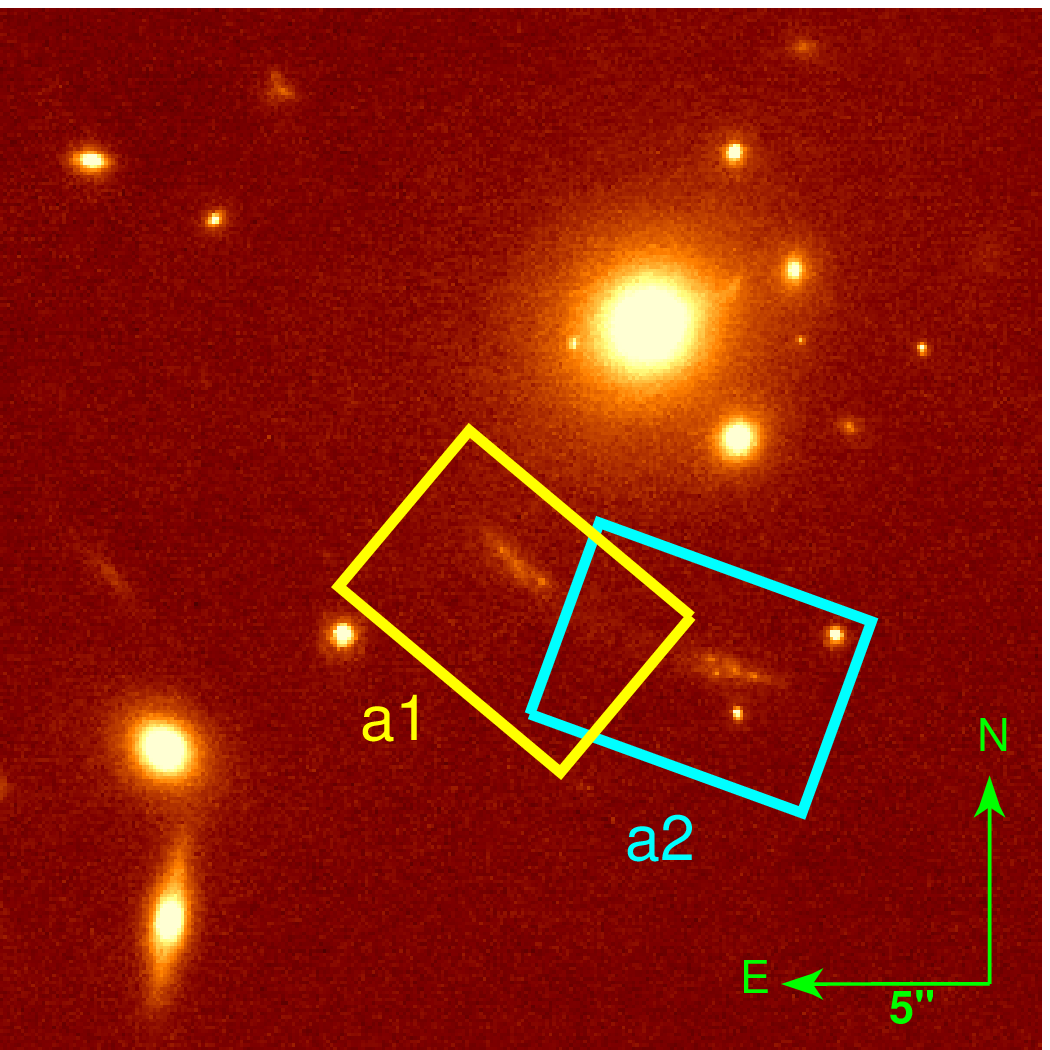}
\caption{\textit{\bf Left:} SDSS color composite image of cswa128, lensed galaxy behind a galaxy group at $z = 0.214$ (red: SDSS z' filter, green: SDSS r' filter, blue: SDSS g' filter). The OSIRIS pointings are indicated by two rectangles, each covering a different multiple image of the lensed galaxy at a mean systemic redshift of $z = 2.225$. Yellow rectangle a1 (t$_{exp}$ = 5.4ks) represents OSIRIS field for previous observations by \protect\citetalias{Leethochawalit16} while our new observations are shown by the cyan rectangle a2 (t$_{exp}$ = 5.4ks). \textit{\bf Right:} Our new NIRC2 Kp band imaging data (t$_{exp}$ = 8.4ks) with OSIRIS pointings overlaid on top. The spatial resolution has improved by a factor of 10 with NIRC2 observations. The length of the compass indicates an angular scale of $5. \!\!^{\prime\prime}0$.}
\label{fig:fig1}
\end{figure*}
\section{Observations and Data Reduction}\label{sec:data}
\subsection{Previous Data}
cswa128 is one of the brightest (r = 19.6) lensed galaxies included in the CASSOWARY survey with a mean systemic redshift of 2.225 derived using H$\alpha$ and [OIII] $\lambda$ 5007 emission lines \citep{Stark13}. Publicly available \textit{gri} imaging data from SDSS DR8 (Figure~\ref{fig:fig1}) had been used previously for creating a lens model for this system. The lensing galaxy group was spectroscopically confirmed to be at $z = 0.214$ \citep{Stark13}. \citetalias{Leethochawalit16} conducted a follow up study using AO aided near-infrared IFU observations with OSIRIS \citep{Larkin06} instrument on  the W. M. Keck I 10 m telescope. The physical location of their OSIRIS field is shown in yellow on the SDSS color composite  in Figure~\ref{fig:fig1}.

\subsection{New Data}
\subsubsection{NIRC2 Imaging data}\label{sec:nirc2data}
To obtain a higher resolution image of the lensed system, cswa128 was observed with the near-infrared Camera (NIRC2) on 2017, September 8 UT using the laser-guide star (LGS) AO system on Keck II (PI: S.Sharma). Observations were taken in wide field mode ($40 \ \!\!^{\prime\prime} \times 40 \ \!\!^{\prime\prime}$, plate scale $0.\!\!^{\prime\prime}04$/pixel) using two filters Kp ($\lambda_{0} = 2.12$  $\mu m, \Delta \lambda = 0.35$  $\mu m$), J ($\lambda_{0} = 1.24$  $\mu m, \Delta \lambda = 0.16$  $\mu m$). Wide field mode was an optimal choice to obtain high resolution and cover the group members in the field of view (FOV) at the same time. A tip-tilt (TT) star with R $\sim$ 15.8 mag was chosen outside the FOV for these observations. Three point dithering ($5 \ \!\!^{\prime\prime}$) pattern was adopted to avoid the noisy fourth quadrant of the detector. The total on-target exposure time was 8.4ks (Kp band) and 4.2ks (J band). J-K$_{p}$ color image was mostly used for selecting the group members in our lens model, however due to relatively less exposure time in J band, we only present Kp band in this paper. Figure~\ref{fig:fig1} shows the NIRC2 FOV for our observation. The spatial resolution of $0.\!\!^{\prime\prime}04$ obtained using NIRC2 allows us to resolve fainter substructures in the lensed galaxy as shown in Figure~\ref{fig:fig3}.

The NIRC2 images were reduced with IRAF using the basic procedures that involved flat-fielding, sky subtraction and distortion correction. Flat-fielding was done by subtracting a stack median of lamp-off from lamp-on dome flats. The process of sky-subtraction was done iteratively. We constructed a sky image for every exposure using a sigma-clipped mean of its neighbouring exposures at different dither positions. After subtracting sky from the respective exposures, images were corrected for geometrical distortion and then stacked together to create an object mask. The objects identified in this mask were rejected for the next round of sky subtraction. Sky subtraction was repeated for a few more iterations till we converged on the final image after combining all the sky-subtracted frames. We used the 2MASS photometry to perform the flux calibration of our NIRC2 image. The astrometry was calibrated using SDSS imaging data by aligning location of the bright stars in the FOV.

\subsubsection{OSIRIS IFU data} \label{sec:fitproc}
We obtained new data using OSIRIS with the natural guide star (NGS) AO system \citep{Wizinowich06} on the a1 multiple image (Figure~\ref{fig:fig1}) of the lensed galaxy. The observation was conducted on 2015, September 22 UT at ($\alpha_{2000}$, $\delta_{2000}$) = (19:58:35.117, +59:50:51.51)(PI:T.Yuan). The target was observed in the Kn2 band to cover the redshifted H$\alpha$ and [N\,{\sc ii}\rm] emission lines. We used a plate scale of $0.\!\!^{\prime\prime}1$ which gave a field of view of $4.5 \times 6.4$ arcsec (Figure~\ref{fig:fig1}). The K$_s$ = 9.8 mag NGS was chosen inside the NGS FOV of OSIRIS. A position angle (P.A.) of 70 deg. was used for the observation. We obtained five individual exposures of 900 s in an ABAB dithering sequence to optimize sky subtraction. The net on-target exposure time was 4.5 ks. The average spectral resolution was R $\sim3000$ which corresponds to a resolution in velocity $\sim 40$ km/s. Centering of the IFU was done through short exposures of the TT star prior to science target exposure. The  AO-corrected PSF was $0.\!\!^{\prime\prime}12$. \\

We use the latest available OSIRIS data reduction pipeline \citep{Larkin06} to reduce the individual exposure frames. Major steps in the reduction pipeline include : dark and bias subtraction, flat-fielding, spectral extraction, wavelength calibration and telluric correction. Adjacent pairs of images were used as reference frames to perform sky subtraction through the IDL code of \cite{Davies07}. The individual exposures were then combined using a $3\sigma$ mean clip to reject cosmic rays and bad lenslets. An A0D type standard star FS147 was used to perform telluric correction and flux calibration. The uncertainty in absolute flux calibration is estimated to be within 20$\%$. The absolute astrometry calibration of the produced datacube was done using the high resolution NIRC2 image with an uncertainty of about 1 spaxel ($\sim 0.\!\!^{\prime\prime}1$). 

\begin{figure*}
\centering
\includegraphics[scale=0.6, trim=1.8cm 0.5cm 0.3cm 1.8cm,clip=true,angle=0]{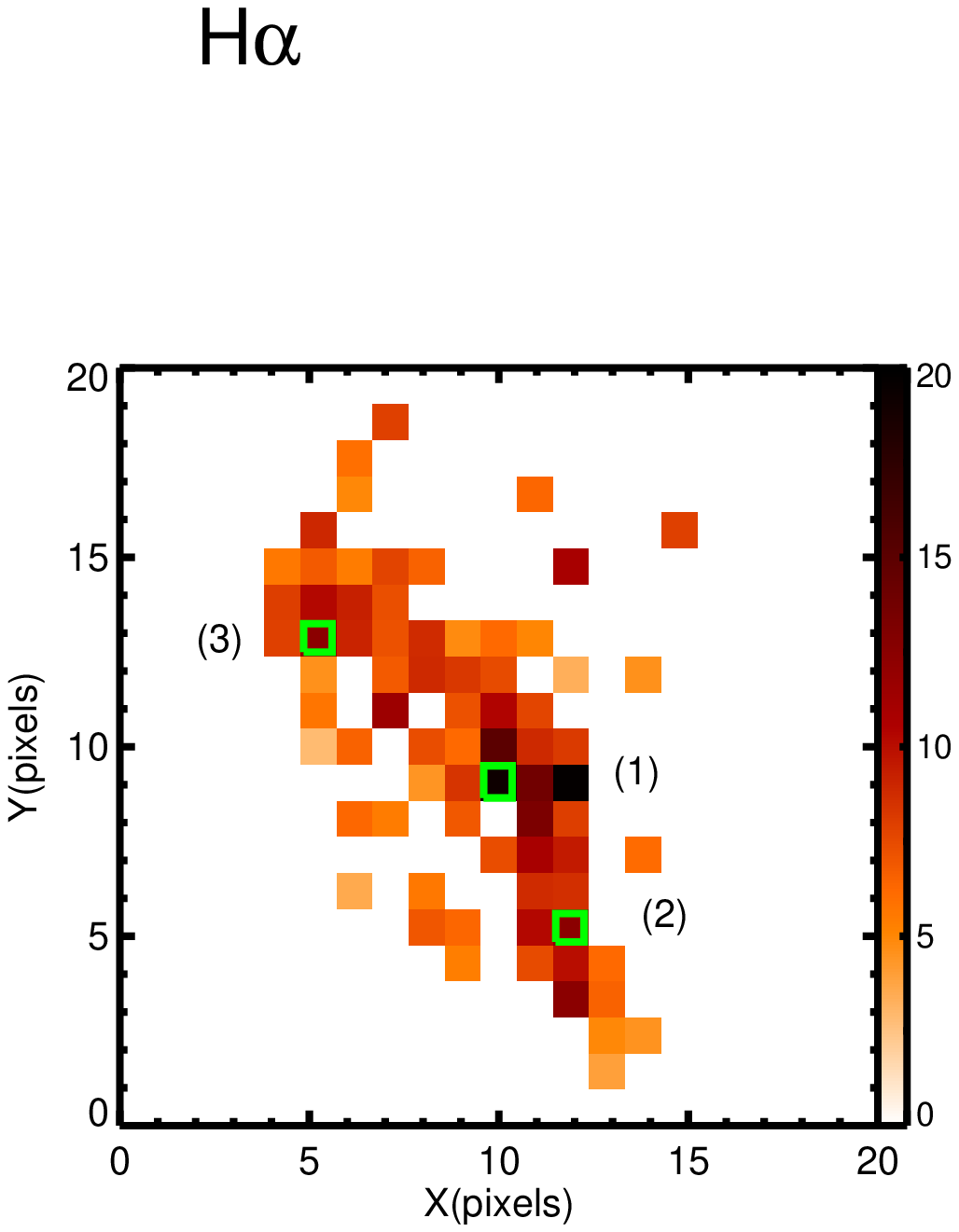}
\includegraphics[scale=0.62, trim=1.3cm 0.5cm 0.0cm 1.8cm,clip=true,angle=0]{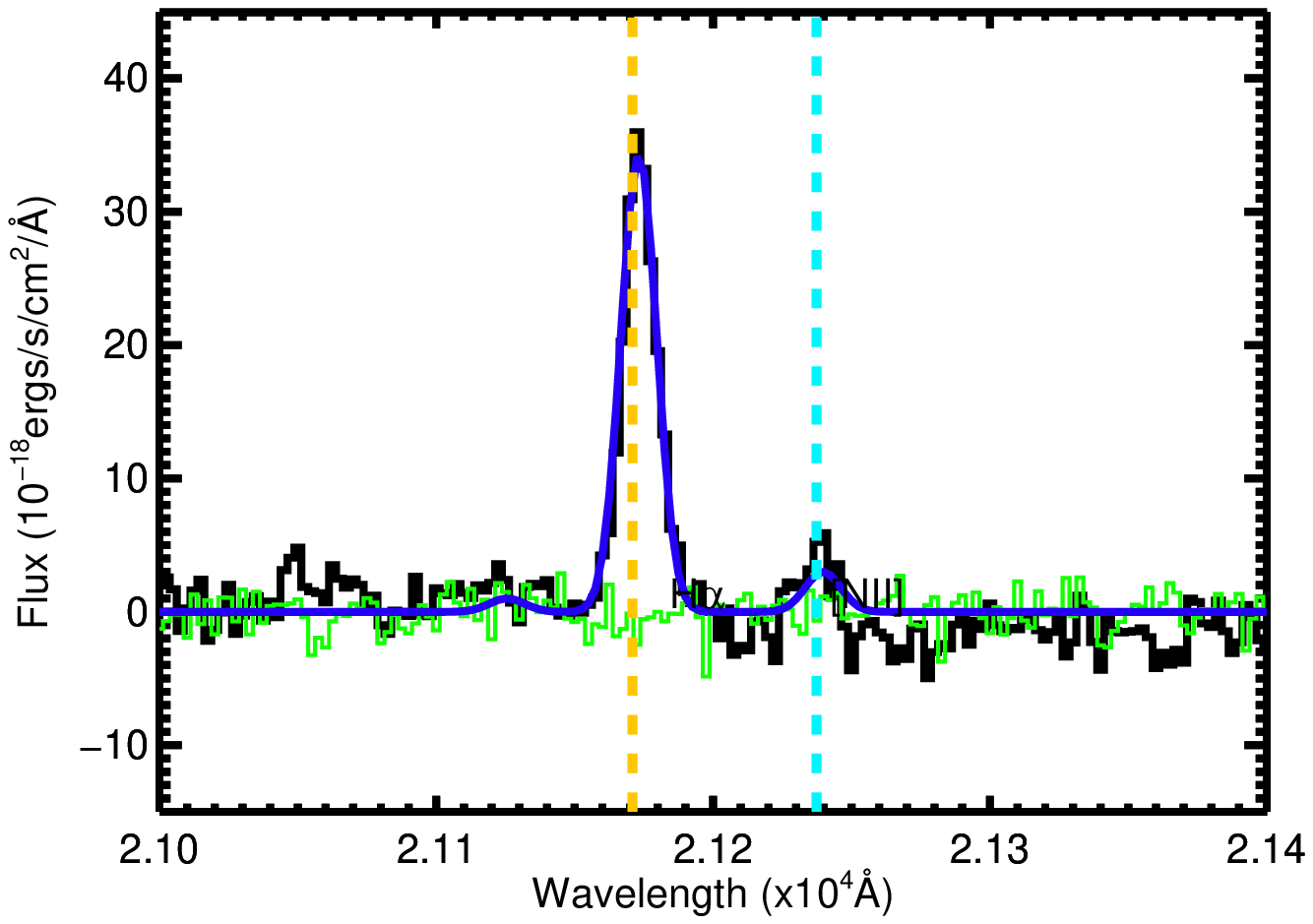}
\includegraphics[scale=0.44, trim=1.4cm 1.0cm 0.75cm 1.8cm,clip=true,angle=0]{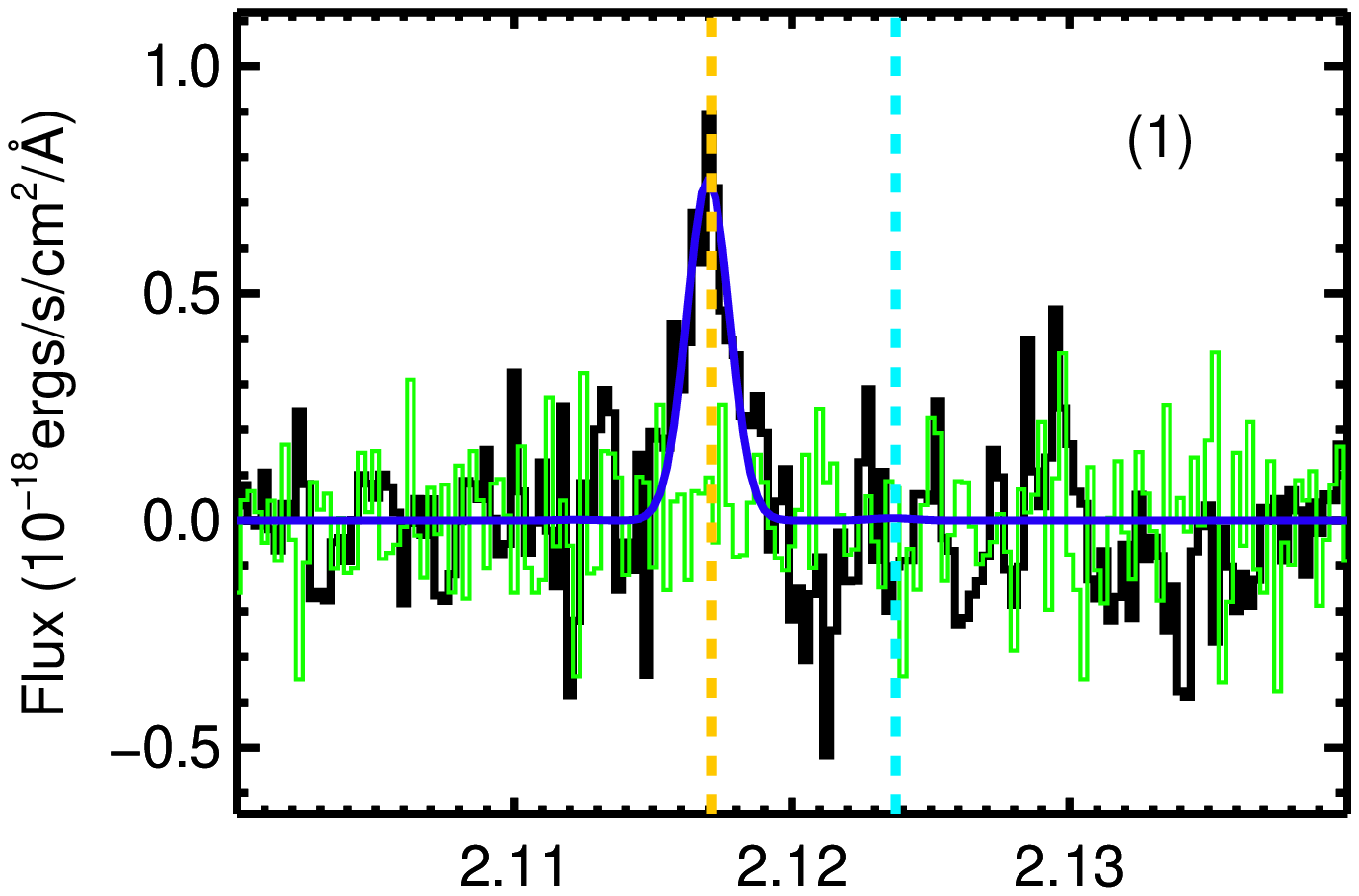}
\includegraphics[scale=0.44, trim=3.00cm 1.0cm 0.75cm 1.8cm,clip=true,angle=0]{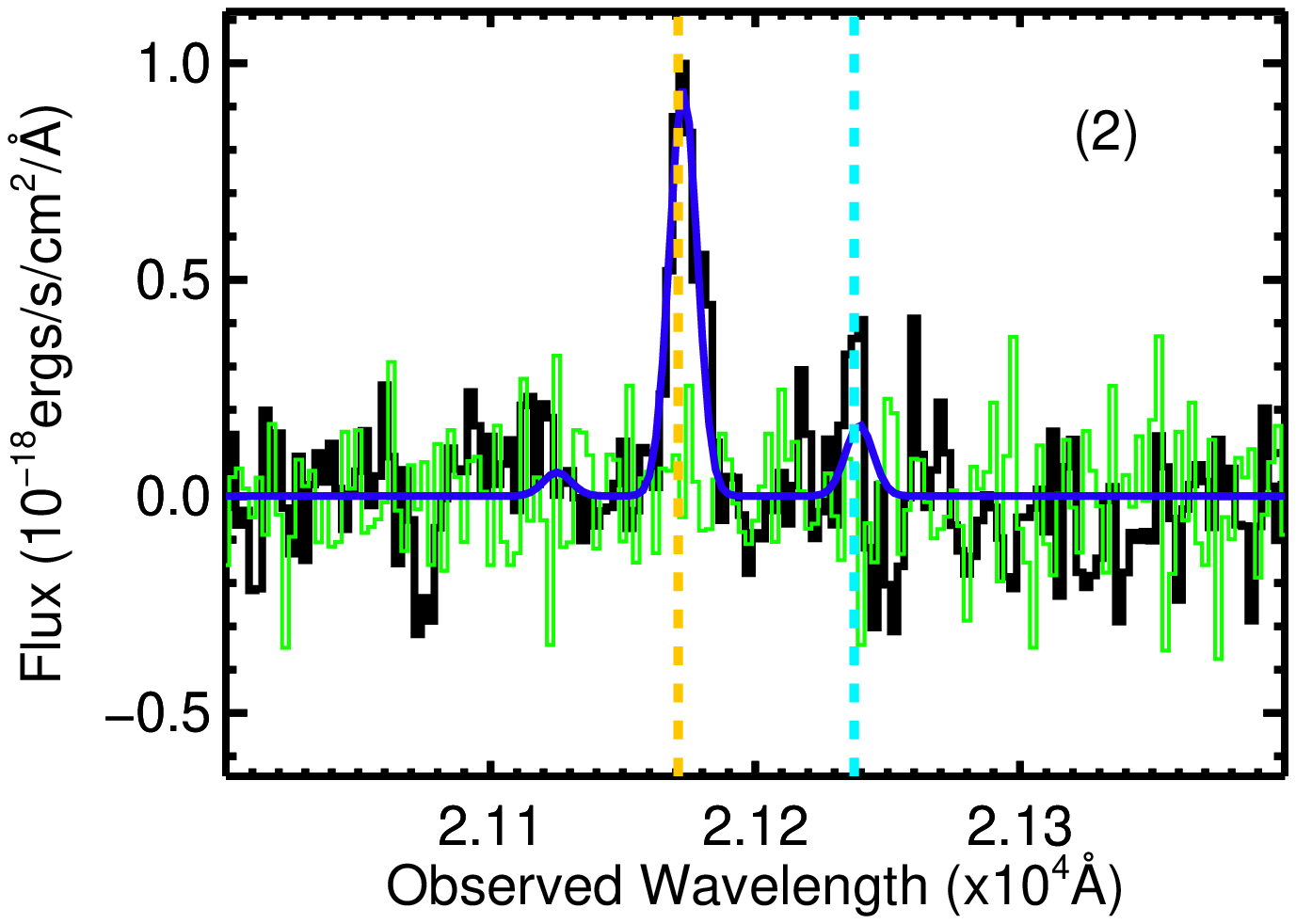}
\includegraphics[scale=0.44, trim=3.00cm 1.0cm 0.0cm 1.8cm,clip=true,angle=0]{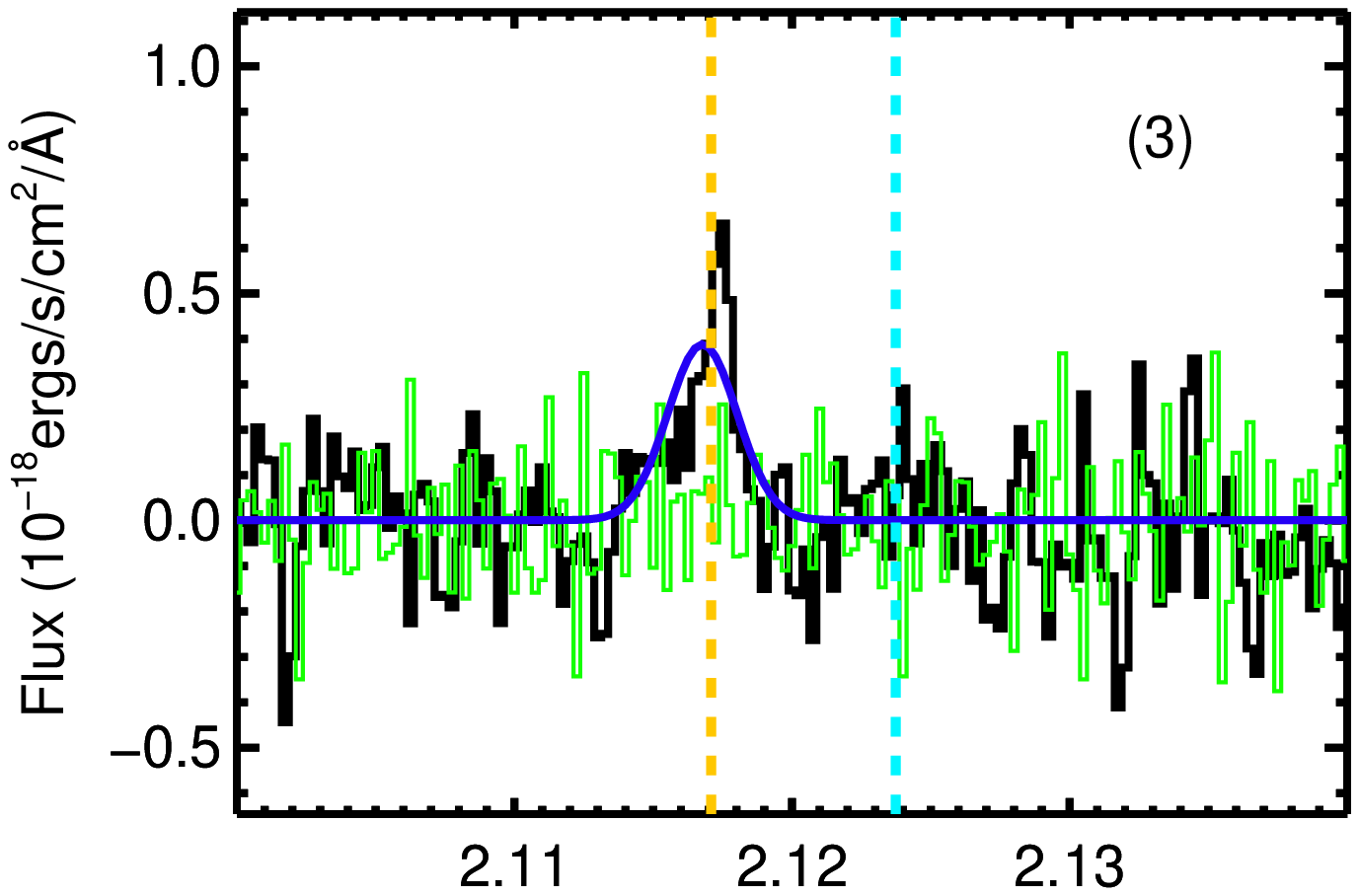}
\caption{ Examples of 1D spectra and H$\alpha$ 2D intensity map from emission line fitting of new OSIRIS spectroscopic data on multiple image a2 of the lensed arc. \textit{\bf Top Left:} H$\alpha$ 2D intensity map in units of $10^{-18}$ erg s$^{-1}$ cm$^2$ arcsec$^{-2}$ in the observed frame. 78 individual spaxels in the image plane achieve H$\alpha$ SNR above 5$\sigma$. 1D spectra of the three green pixels in the 2D map are shown in the bottom three panels. \textit{\bf Top Right:} Integrated 1-D spectrum in the vicinity of H$\alpha$  and [N\,{\sc ii}\rm] lines  obtained after summing all the IFU pixels with H$\alpha$ detections above 5$\sigma$. The flux density is in units of $10^{-18}$ erg s$^{-1}$ cm$^2 \angstrom^{-1}$. We achieve a SNR of 25 on H$\alpha$ and 5 on [N\,{\sc ii}\rm]. Note that the weaker [N\,{\sc ii}\rm] $\lambda 6548$ line is fixed to be 1/3 of the [N\,{\sc ii}\rm] $\lambda 6584$ line. The negative level of continuum is due to the its over-subtraction especially in the red part of the spectrum. \textit{\bf Bottom:} Panels shows the spectrum of individual spaxels. Yellow and Cyan vertical lines in Panels 1-4 indicate the location of redshifted H$\alpha$ and red [N\,{\sc ii}\rm] emission lines at $z = 2.225$ respectively. The raw spectrum is plotted in black; RMS of sky residuals in green and gaussian fits to the three lines are shown in blue. }
\label{fig:fig2}
\end{figure*}

The critical analysis for this work is the extraction of emission line maps. We fitted gaussian profiles simultaneously to the three emission lines [N\,{\sc ii}\rm] $\lambda6548, 6583$ and H$\alpha$ after constraining the ratio of [N\,{\sc ii}\rm] $\lambda6548$ and [N\,{\sc ii}\rm] $\lambda6583$ to the theoretical value of 0.33 as described in Osterbrock (1989). We also constrained the mean and velocity width of [N\,{\sc ii}\rm] $\lambda6548, 6583$ lines to be the same as H$\alpha$ $\lambda6563$. Fitting of emission lines was initially done for every spaxel in the image and source plane using a weighted $\chi^{2}$ minimization procedure. Since the wavelength range covering [N\,{\sc ii}\rm] and H$\alpha$ lines was not dominated by the atmospheric OH lines, we followed the weighting procedure outlined in \citetalias{Leethochawalit16}. For a given wavelength, the fitted spectra were weighted by a variance spectrum calculated using an emission line free region of the data cube. The generated 2D map from the emission line fitting had 78 spaxels with  H$\alpha$ SNR $\geq 5$ and 15 spaxels with [N\,{\sc ii}\rm] SNR $\geq 3$ (Figure~\ref{fig:fig2}). We also present the coadded spectrum of all spaxels for comparison in Figure~\ref{fig:fig2}. Moreover, we notice that the spatial averaging of $3 \times 3$ surrounding spaxels fails to improve the SNR significantly. We present an adaptive binning technique using our new lens model to achieve higher SNR. The details of the method are discussed in Section ~\ref{sec:analysis}. 

\section{Lens Modeling}\label{sec:lensmodel}
The previous mass model for the galaxy group of cswa128 was constructed using the procedure described in \cite{Auger13}. The \textit{gri} SDSS imaging data was used to fit a single Sersic profile to each of the lensing galaxy and the background source galaxy using a $\chi^{2}$ optimization procedure. In this work, we use high resolution NIRC2 imaging data to obtain a well constrained lensing mass model. By finding additional substructures in observed images, we build a more robust model for this system. In this section, we describe our parametric approach to construct a strong lensing mass model for this system. 

\subsection{Methodology}
We use the publicly available software {\lenstool}\footnote{http://projects.lam.fr/projects/lenstool} to reconstruct in detail the mass distribution of cswa128. The modeling software package LENSTOOL \citep{Kneib93, Jullo07} adopts a parametric approach to effectively model the mass distribution of lenses. It has been successfully used to model various lensed systems and has reliably reproduced the observations with best image plane accuracies \citep{Jauzac15, Limousin16, Lagattuta17, Mahler18}. A detailed reconstruction of the lensed galaxy involves the lensing model to capture both the \textit{large} scale structure: caused by the potential of the host halo of the lens (galaxy group) and the \textit{small} scale substructure associated with the locations of member galaxies within the galaxy group. These small scale potentials influence the distortions in lensing potential locally and are constrained using a light traces mass (LTM) approach by {\lenstool}. A Bayesian Monte Carlo Markov Chain (MCMC) minimization procedure in {\lenstool} \citep[details about the procedure can be found in][]{Jullo09} was used to efficiently optimize the free parameters of the system.
 
A truncated Dual Pseudo-Isothermal Elliptical mass distribution \citep[hereafter dPIE;][]{DPIE07} was used to model the mass components at both small and large scales. The dPIE profile is characterized by seven parameters: central position (x,y), orientation angle $\theta$, ellipticity $e$, central velocity dispersion $\sigma$, two characteristic radii: r$_{core}$ and r$_{cut}$. We model the group scale halo with a single dPIE profile and allow all its parameters except r$_{cut}$  to vary freely with uniform priors. Since r$_{cut}$ is unconstrained by strong lensing data, it is fixed to a value of 200 kpc, typical for group scale galaxy lenses. 
\begin{figure*}
\centering
\includegraphics[scale=0.6,clip=true, angle=0]{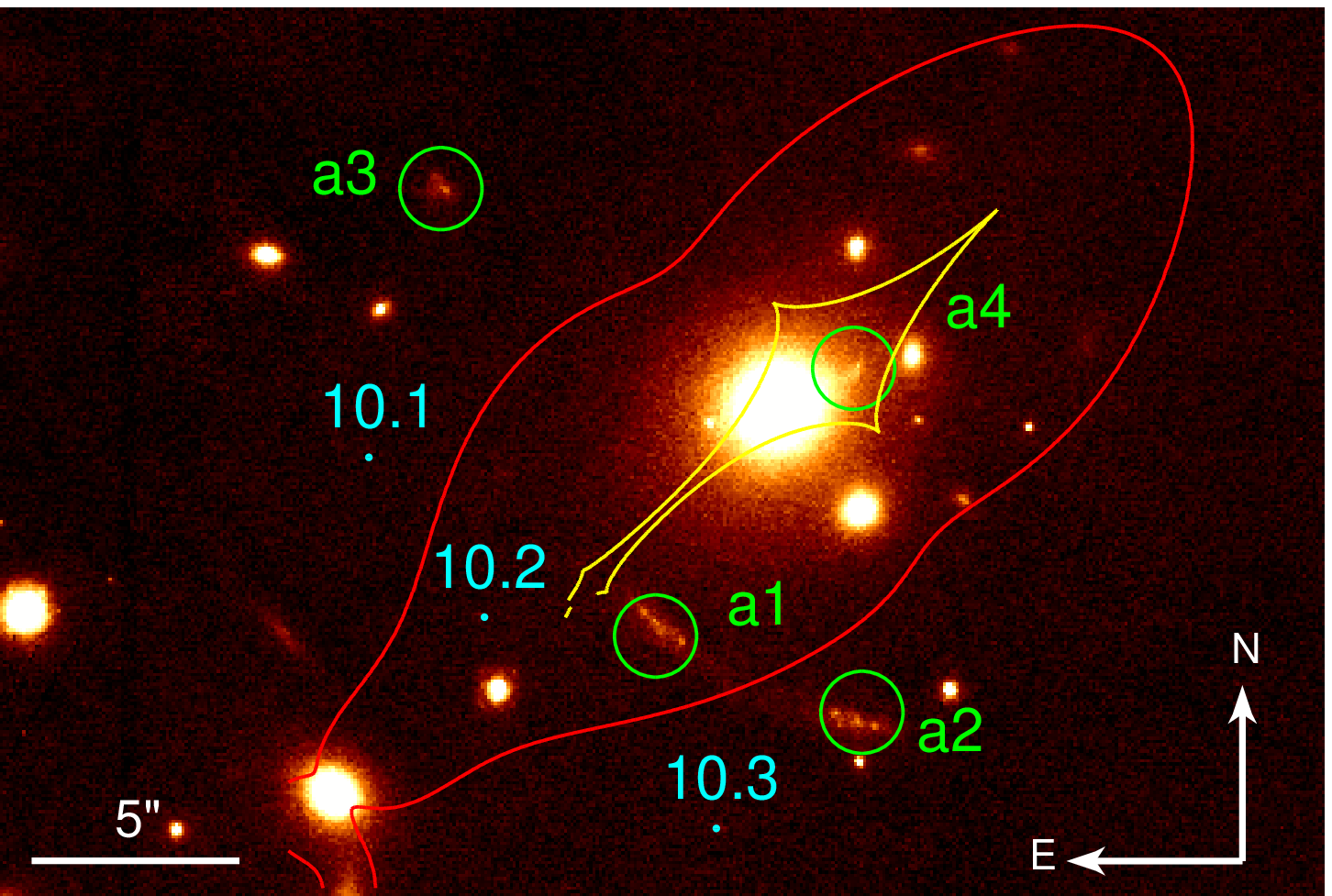}
\includegraphics[scale=0.6,clip=true, angle=0]{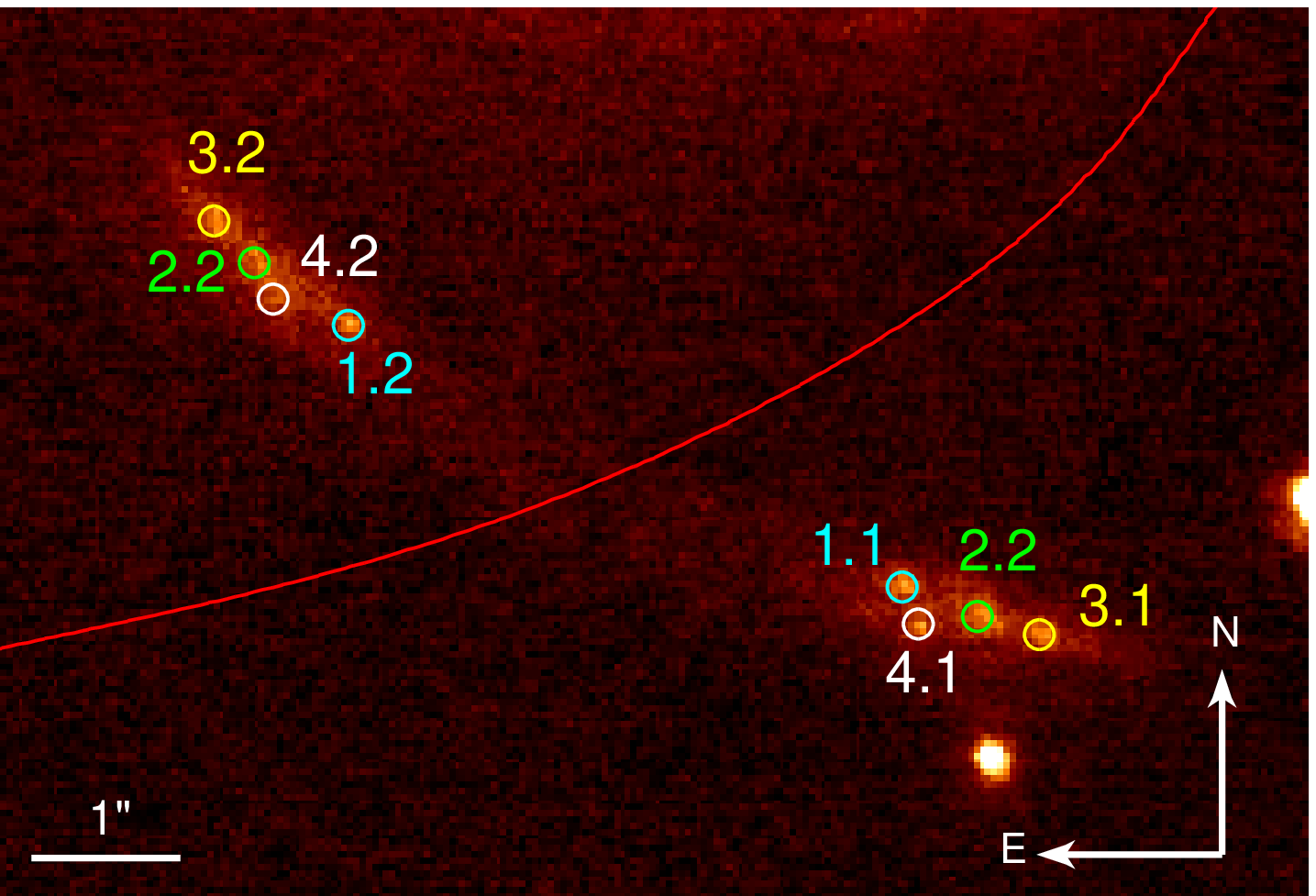}
\caption{Our lens model of CSWA128. \textit{\bf Left:} NIRC2 imaging of CSWA128 in Kp band. Labeled in green circles (a1, a2, a3, a4) are the different multiple images of the lensed galaxy at  $z = 2.225$. IFU data for a1 and a2 confirm their mean redshifts at 2.225 whereas a3 and a4 have been identified on the basis of similar morphology using NIRC2 image and initial predictions from our lens model. The critical lines at $z = 2.225$ are shown in red and the corresponding caustics in yellow. System 10 has been identified as a separate system and included in our model with redshift as a free parameter. \textit{\bf Right:} Close-up image of the lensed arc to show smaller substructures used as constraints in our mass model. Each color corresponds to a different set of constraint, matched across in the two multiple images.}
\label{fig:fig3}
\end{figure*}

To limit the parameter space, group members close to the arc within the strong lensing regime were included in the lens model. Individual galaxy clumps were also modelled with a dPIE profile, after fixing the associated geometrical parameters (x, y, $e$, $\theta$) by running SExtractor \citep{Bertin96} on the NIRC2 image presented in section~\ref{sec:nirc2data}. Other parameters ($\sigma$, r$_{cut}$ ) were scaled on the basis of K-band luminosity of galaxies (L$_{K}$), relative to a L$_{K}^{*}$ galaxy using a constant mass-luminosity scaling relation \citep{Faber76} as described by the following equations \cite[see also discussion in][]{Richard10} :
\begin{eqnarray}
&r_{core} = r_{core}^{*}( L_{K} / L_{K}^{*})^{1/2} , &\\ 
&r_{cut} = r_{cut}^{*}( L_{K} / L_{K}^{*})^{1/2} , &     \\ 
&\sigma_{0} = \sigma_{0}^{*}( L_{K} / L_{K}^{*})^{1/4}& 
\end{eqnarray}
 
We fixed r$_{cut}^{\star}$ to 50 kpc but allowed $\sigma_{0}^{\star}$ to vary with a gaussian prior of mean $\mu = 158$ km/s and width $\sigma = 27$ km/s following the observational results of \cite{Bernardi03} on $\sigma^{\star}$. Since r$_{core}$ is typically small and does not have any significant effect on modeling results \citep[][]{Covone06, DPIE07, Limousin07}, it was fixed to 0.15 kpc. We did not include the BCG in the scaling relations because it might have a significantly different mass to light ratio as a result of its distinct formation history. The BCG was therefore modelled as a potential with $\sigma$ as the only free parameter. In total we had eight free parameters in our model all of which have broad uniform priors. 

We make use of the spectroscopic information and imaging data to identify different multiple images of the lensed galaxy. The counterparts a3 and a4 to the lensed arc (Figure~\ref{fig:fig3}) have been identified on the basis of their image configuration and similar morphology to a1 and a2. Furthermore, high resolution NIRC2 image allows us to uniquely identify various features  (Systems 1-4 in Figure~\ref{fig:fig3}) within the lensed arc and match them across multiple images a1 and a2. These knots are used as constraints in our lens model (See Table~\ref{tb:tb1}). The lower magnification of a3, a4 does not allow us to resolve similar features even with the high resolution data.

The model is then computed as an iterative optimisation process. The constraints come from the location of multiply imaged systems, where a positional uncertainty of $0.\!\!^{\prime\prime}1$ is considered. During the optimisation, LENSTOOL produces different MCMC realizations minimising the RMS between the observed and predicted positions of all images. An initial set of constraints is fed into LENSTOOL and a basic lens model is created, where many parameters are degenerate with each other. We then use it to test different assumptions on the multiply-imaged pairs and add new constraints to improve the optimisation of the model and break any degeneracies. For example, multiple image system 4 in Figure~\ref{fig:fig3} was included as a constraint in the later stages of modeling with the help of our early lens model which confirmed its identification. We also include the location of a secondary lensed system (labelled 10 in Figure~\ref{fig:fig3}) as a constraint in our lens model and optimize its redshift during the lens modeling. For more details on such a modeling procedure, please refer to \cite{Limousin07}, \cite{Richard14} and \cite{Verdugo11}.

\begin{figure*}
\centering
\begin{subfigure}[a]{0.95\textwidth}
\includegraphics[scale = 0.65,trim=0.0cm 0.0cm 0.0cm 0.0cm,clip=true, angle=0]{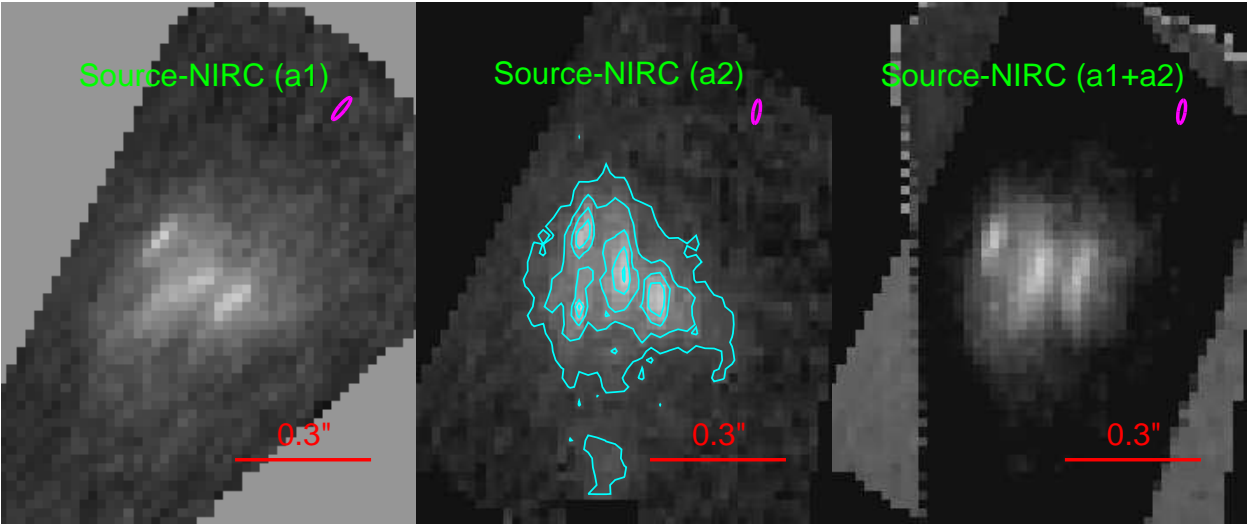}
\caption{Source Reconstruction of NIRC imaging data for each of the multiple images (a1, a2) of the lensed arc and their rendition into one frame using the best-fit lens model. Subsampling of 2 in source plane and 10 in the image plane is used to obtain a source-plane pixel size of $0.\!\!^{\prime\prime}02$. The reconstructions are astrometrically aligned with the IFU - H$\alpha$ source plane images shown in Figure~\ref{fig4:sub2}. Cyan contours highlight the different surface brightness levels of the source reconstructed a2 image of the lensed arc. Being the most magnified image of the galaxy, we use its contours to define the stellar emission from the galaxy. The last panel shows the combined source reconstructed image after manually adjusting the differences in position and orientation of a1 to match that of a2.}
\label{fig4:sub1}
\end{subfigure}
\begin{subfigure}[b]{0.95\textwidth}
\includegraphics[scale = 0.75,trim=0.0cm 0.0cm 0.0cm 0.0cm,clip=true, angle=0]{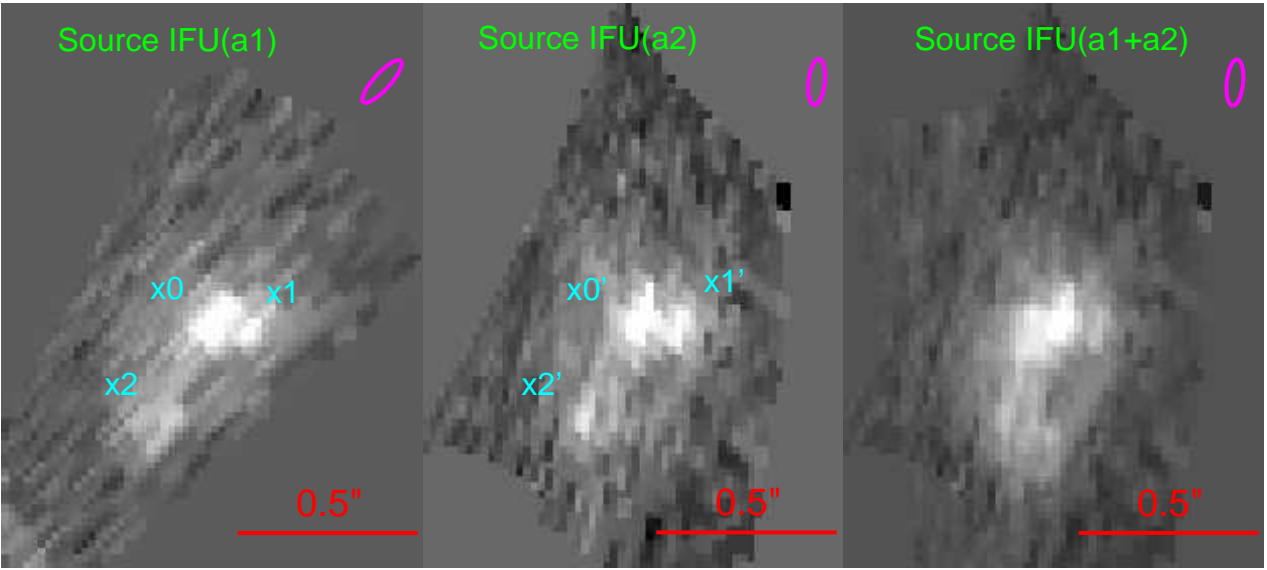}
\caption{Source Reconstruction of IFU data for each of the multiple images (a1, a2) of the lensed arc and their rendition into one frame using the best-fit lens model. About 6 datacube slices close to the redshifted H${\alpha}$ wavelength are stacked to form a co-added 2D image and then sent to the source plane. Subsampling of 5 in source plane and 10 in the image plane is used to obtain a source-plane pixel size of $0.\!\!^{\prime\prime}02$. Clumps marked in cyan are matched across in the first two panels showing two independent reconstructions of the galaxy. The third panel shows the coadded image after manually adjusting the differences in position and orientation of a1 to match that of a2.}
\label{fig4:sub2}
\end{subfigure}
\caption{An example of source reconstructed stellar continuum from NIRC2 observations and H$\alpha$ brightness distribution from IFU data for each of the magnified images (a1, a2) of the lensed arc and their rendition into a single frame to highlight the disadvantages of source-plane PSF. Magenta ellipses represent source-plane PSF's. As clear from the last image in both the panels, final merged image does not allow a significant improvement in SNR of different features labelled in cyan over individual reconstructions.}
\label{fig:fig4}
\end{figure*}

It is apparent that SDSS data lacks the required resolution and depth to identify different counterparts and arc substructures. This limits the accuracy of the model to interpret IFU observations. Using our modified model, we reach an astrometric precision of $0.\!\!^{\prime\prime}06$ in the predicted image positions. The mean flux magnification (measured by averaging the magnification for individual clumps constituting the arc) across a1  is $\mu_{1} = 9.09 \pm 2.86$ and a2 is $\mu_{2} = 11.56 \pm 5.98$.

\begin{table}
\centering
\caption{Constraints used in modeling cswa128, labelled in Figure~\ref{fig:fig3}.}
\begin{tabular}{lccc}
\hline
\hline
Arc ID   &   R.A J2000  &  Dec. J2000 & $z$ \\
             & (deg)  & (deg) &  \\ 
\hline
1.1   &   299.6462 & 59.8477  & 2.225 \tablefootnote{The redshifts of systems 1 - 4 are fixed to the spectroscopic redshift.}\\  
1.2   &   299.6483 & 59.8481  & ... \\
\hline
2.1   &   299.6459 & 59.8476  & ... \\  
2.2   &   299.6486 & 59.8483  & ... \\
2.3   &   299.6515 & 59.8512  & ... \\  
2.4   &   299.6459 & 59.8499  & ... \\
\hline
3.1   &   299.6457 &  59.8476 & ... \\  
3.2   &   299.6488 & 59.8484  & ... \\
\hline
4.1   &   299.6461 & 59.8476  & ... \\  
4.2   &   299.6485 & 59.8482  & ...\\
4.3   &   299.6515 & 59.8512  & ... \\
\hline
10.1  &  299.6524 & 59.8494  & 2.900 $\pm$ 0.25 \tablefootnote{The redshift of system 10 is optimized by the lens model.}\\
10.2  &  299.6509 & 59.8483  & ... \\
10.3  &  299.6478 & 59.8469  & ... \\
\hline
\end{tabular}
\label{tb:tb1}
\end{table}%

\section{Lensing Analysis}\label{sec:analysis}
Lensing mass models play an important role in deriving the source plane properties of the background lensed galaxies. We derive the image to source plane mapping using the lens model and correct IFU observations for lensing distortion. The focus of this study is mainly threefold: examine the source in terms of its morphology;  derive kinematical patterns through velocity gradients and investigate chemical gradient through emission line ratios in the galaxy. In order to optimize the SNR of emission lines in the source plane, we merge the two independent observations, each from a different multiple image of the galaxy. The derived quantities from the coadded distribution undoubtedly provide a more reliable means to understand the physical conditions within the galaxy. 

Using the lens model, we employ a forward modeling approach to combine the matching spaxels (which map to the same source position) from the two IFU datacubes. Then we fit the emission lines for the combined data to derive line fluxes, velocity, and velocity dispersion in the source plane. Moreover, since our technique is adapted to lens-model magnification, we are able to achieve enhanced H$\alpha$ SNR at a resolution of $\sim$170 pc in highly magnified subregions of the source. 

\subsection{Traditional Ray Tracing Approach: Direct Image to Source Plane Mapping} 
We first use the \textit{cleanlens} task in {\lenstool}\footnote{https://projets.lam.fr/projects/lenstool/wiki/LenstoolManual} to convert the image plane maps to their corresponding source plane distributions. For each position in the image plane, this task computes the corresponding position in the source plane using the ray-tracing provided by our best-fit lensing model. We preserve the surface brightness to obtain the intrinsic source plane fluxes. The source-plane reconstructions are shown in Figure~\ref{fig:fig4}.

This method breaks down as we proceed further with combining the individual reconstructions to obtain the overall source -plane profile. Because lensing preferentially shears the galaxy along a specific direction, there is a source-plane PSF associated with every reconstruction. The size of the source-plane PSF provides a reliable measure of the  physical resolution we can achieve in the source plane. The orientation of the source-plane PSF is defined by the direction of lensing amplification. Reconstruction of different lensed images on the source plane will vary in their source-plane PSF's, both in size and orientation, making the co-adding a non-trivial problem. Traditionally, multiply lensed arcs were combined by a simple averaging after correcting for the offsets in position and PSF orientation \citep{Yuan17, Livermore12, Sharon12, Colley96}. However, as illustrated in Figure~\ref{fig:fig4}, a simple averaging of two images a1 and a2 will not provide an ideal improvement in the achievable SNR in the source plane, thus being of no significantly better use than the individual reconstructions. 

\subsection{Forward modeling Approach: Source to Image Plane Mapping}\label{sec:forward_approach}
In this section, we describe our forward modelling approach to combine the IFU observations of two images (a1, a2) of cswa128. We ray-trace every spaxel from source to image plane at the location of two multiple images and convolve them with their respective PSFs. Following this technique, we generate a co-added 1-D spectrum for every source-plane spaxel by summing together the corresponding ray-traced spaxels convolved with the respective PSF's in the respective image-plane datacubes. This method not only circumvents the problem of combining different source plane PSF's but also provides a clever way to bin the image spaxels to maximize SNR in the source plane.  

\begin{figure*}
\centering
\begin{subfigure}[a]{0.95\textwidth}
\includegraphics[scale=0.45,angle=0,trim=0.0cm 0.0cm 0.7cm 0.0cm,clip=true]{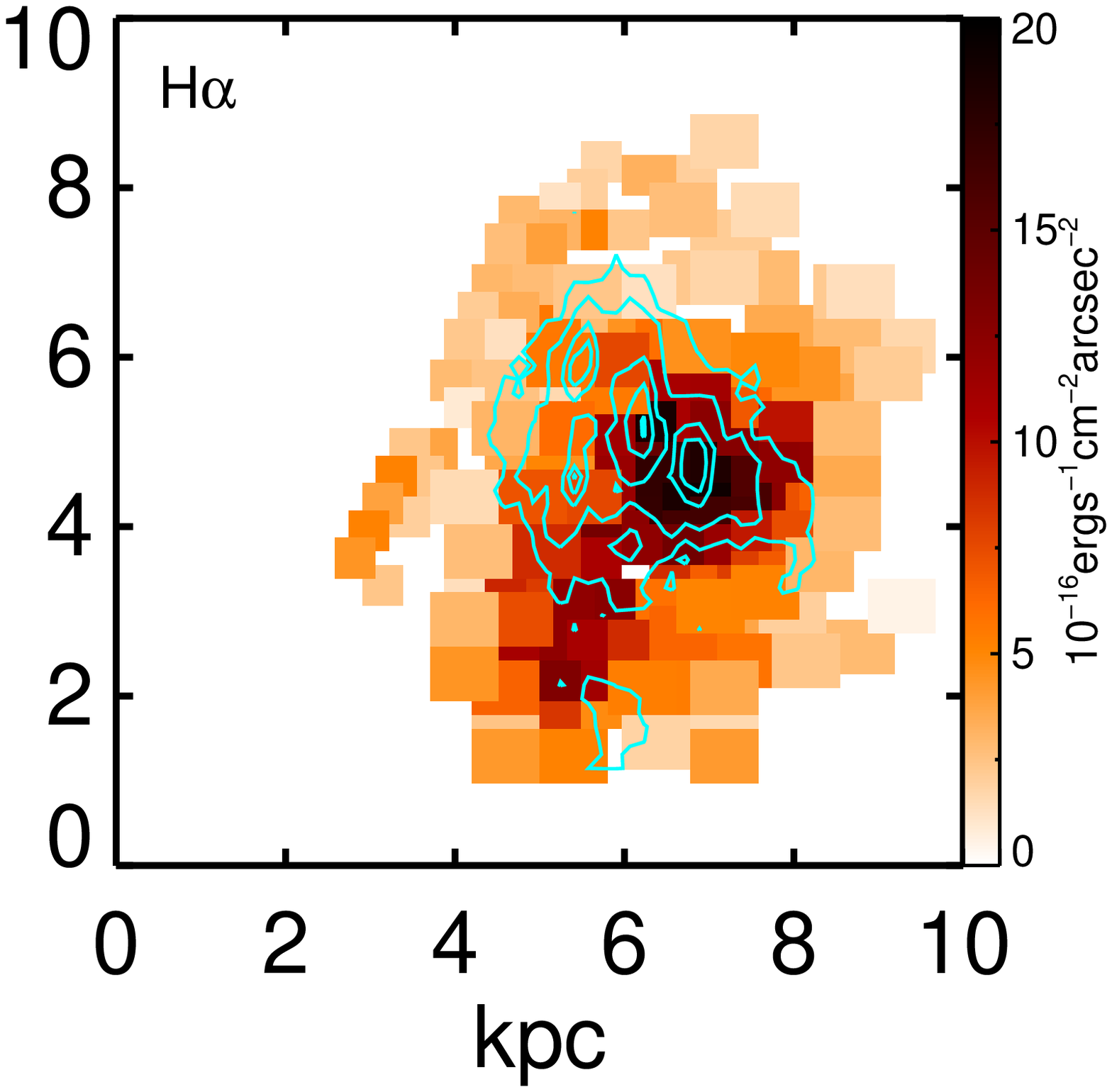}
\includegraphics[scale=0.455,trim=1.0cm 0.0cm 0.7cm 0.0cm,angle=0]{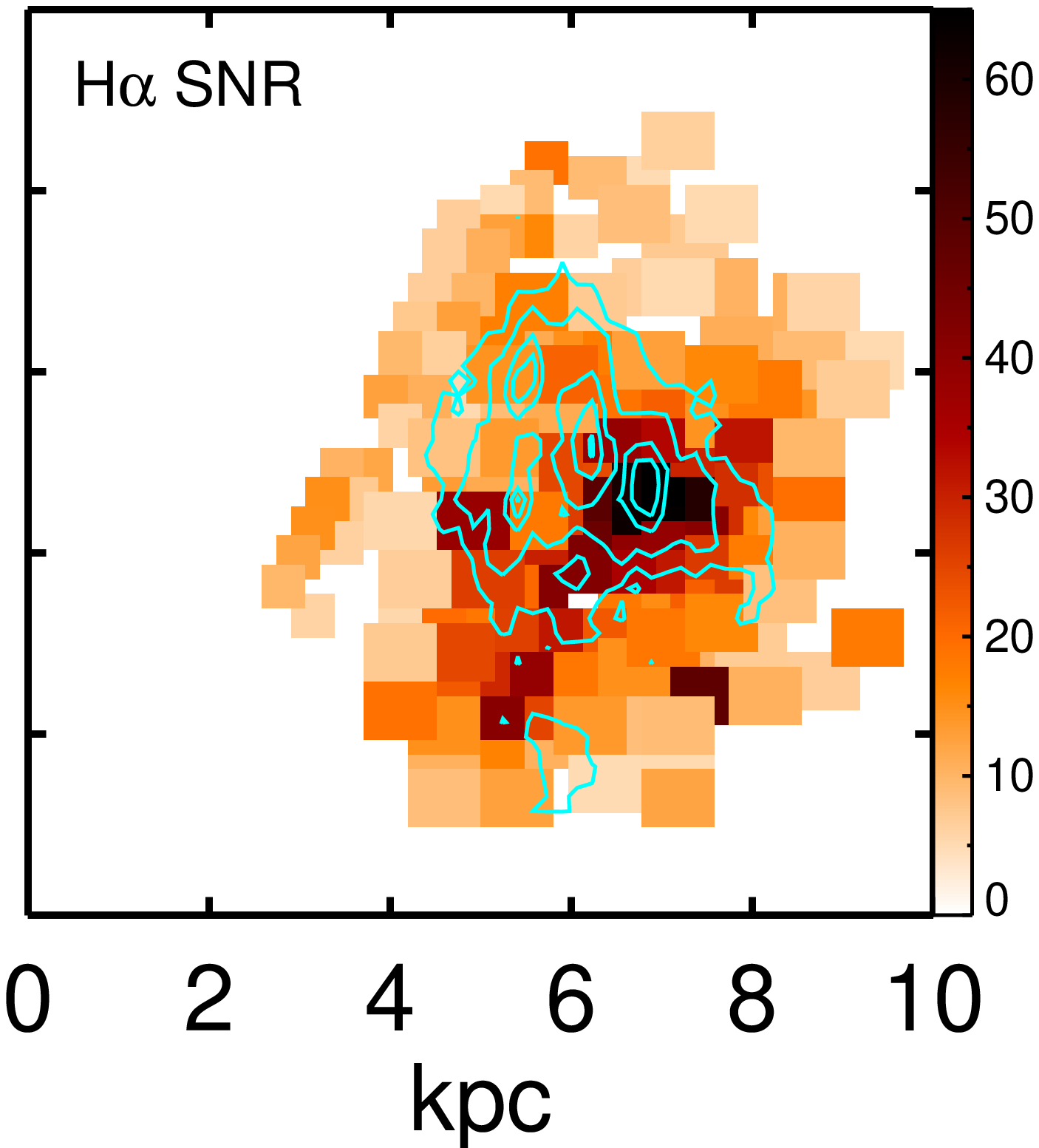}
\label{fig6:sub1} 
 \caption{Derived 2D H$\alpha$ intensity in units of $10^{-16}$ erg s$^{-1}$ cm$^2$ arcsec$^{-2}$ and SNR maps for combined source plane data using our forward-adaptive co-adding technique. Contours are the same as Figure~\ref{fig:fig4}.}
\end{subfigure}
\begin{subfigure}[b]{0.95\textwidth}
\includegraphics[scale=0.45,angle=0,trim=0.0cm 0.0cm 0.7cm 0.0cm,clip=true]{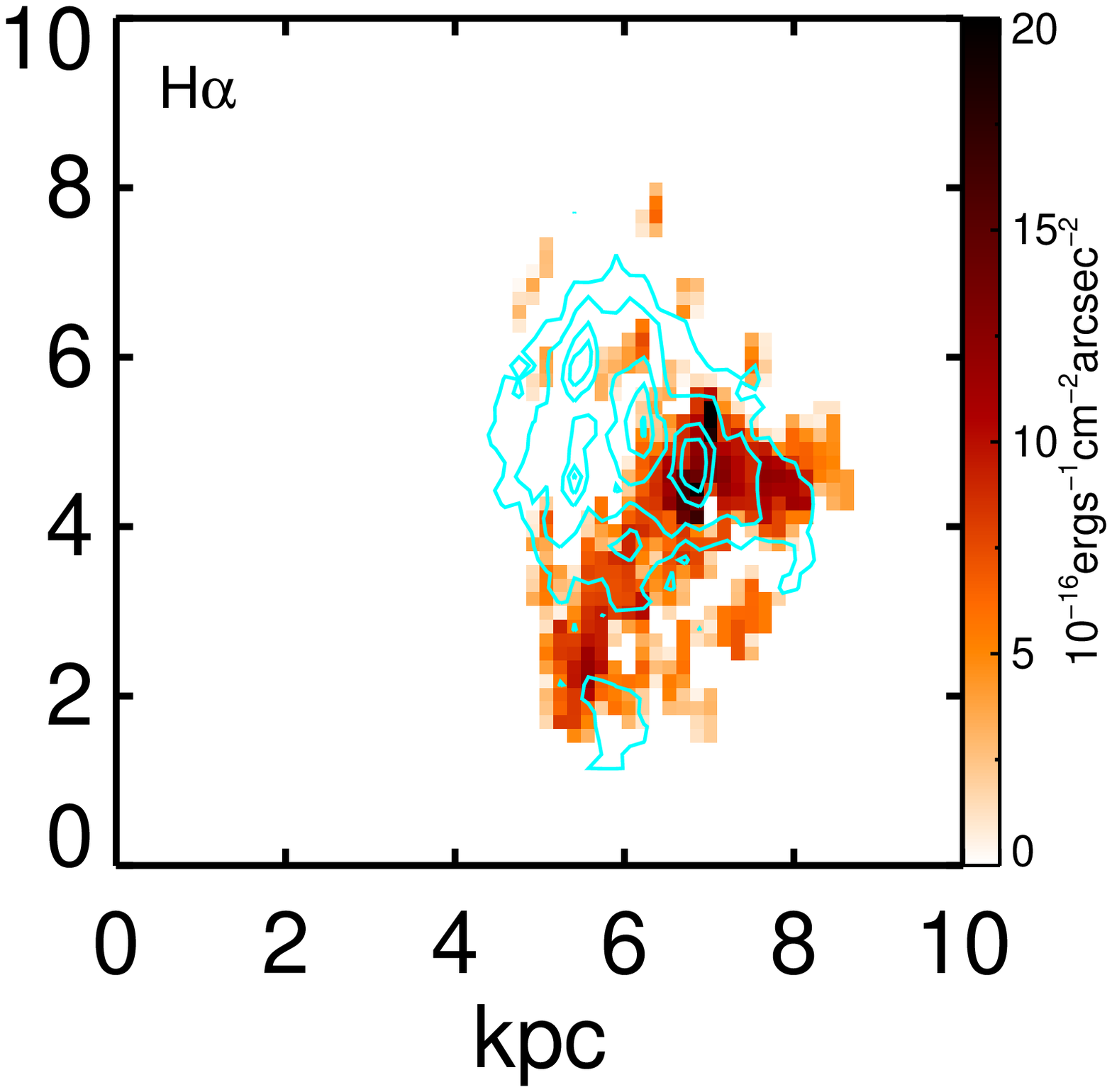}
\includegraphics[scale=0.455,angle=0,trim=1.0cm 0.0cm 0.7cm 0.0cm,clip=true]{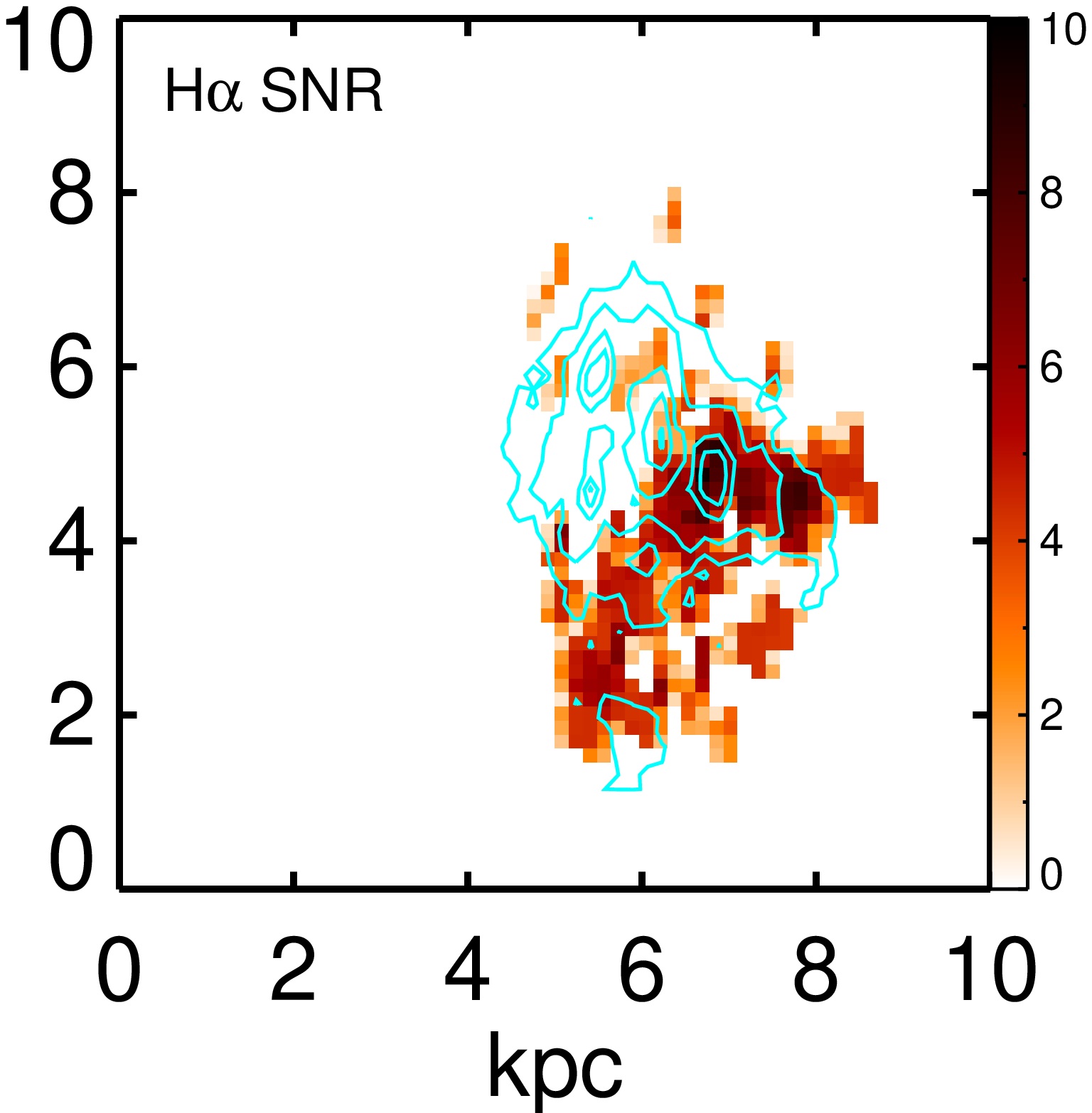}
 \label{fig6:sub2}
 \caption{Source reconstructed 2D H$\alpha$ intensity in units of $10^{-16}$ erg s$^{-1}$ cm$^2$ arcsec$^{-2}$ and SNR maps for multiple image a2 using traditional ray-tracing methodology. Contours are the same as Figure~\ref{fig:fig4}.}
\end{subfigure}
\caption{Comparing the 2D H$\alpha$ intensity and SNR maps using (a) our forward coadded source-reconstruction with (b) the traditional ray tracing approach of multiple image a2 of cswa128. We obtain an SNR improvement of 6 $\times$ using our new approach. Figure also shows the difference in morphologies of the ionized gas emissions (as traced by H$\alpha$) and the stellar continuum (from NIRC2 observations) shown by cyan contours.}
 \label{fig:fig5}
\end{figure*}

The overall procedure is outlined in the following steps:-
\begin{enumerate}
\item\label{step1}We use the \textit{cleanlens} task of {\lenstool} with a grid over-sampling of \textit{ech} = 10 on the image and \textit{sech} = 5 on the source plane to create a template source plane frame. The source plane pixel-scale after over-sampling is $0.\!\!^{\prime\prime}02$. This template frame is only created for determining the source plane region that corresponds to the FOV of observations in the image plane. 
\item Next, we create a series of source plane masks, each containing certain number of selected pixels. These masks, when added together cover adequately the entire region of the template source frame created in the previous step. The number of highlighted pixels in each mask depend on their location with respect to the caustic in the source plane. Because the source-plane pixels lying close to the caustic are highly magnified, they map to a larger part of image area. In such regions, we define masks of individual pixels whereas for regions with lower magnification, we increase the number of selected pixels. In low $\mu$ - high SNR regions of the data cube we resort to single pixel masks.  
\item Each of the source plane masks are then sent to the image plane using {\lenstool}. The simulated image plane masks for each multiple image involve convolution by image PSF so that they can be directly compared with the observed images. We match the simulated masks with the observed datacubes to determine the image-plane spaxels ray traced from each of the source-plane masks.
\item\label{step4}This set of image plane pixels from the two IFU datacubes are then averaged together for each wavelength to  create a 1D spectrum associated with every source plane mask. 
\item\label{step5} Emission lines, H$\alpha$ and [N\,{\sc ii}\rm] are fit simultaneously for each 1D spectra following the same fitting procedure as described in section~\ref{sec:fitproc}. We then manually inspect each fitted spectrum to reject spurious detections. We also note that the fitted parameters don't change significantly with a range of different initial conditions for the source plane spaxels in the fitting routine. We set a SNR cut of 5$\sigma$ on H$\alpha$ and 3$\sigma$ on [N\,{\sc ii}\rm] emission lines. If the achieved H$\alpha$ SNR < 5$\sigma$ as a result of our fitting routine, we increase the size of our masks and repeat steps~\ref{step1}-~\ref{step5}. For source-plane pixels selected in more than one mask, we choose the fitting result with better H$\alpha$ SNR.
\end{enumerate} 

\begin{figure*}
\centering
\includegraphics[scale=0.35,angle=0]{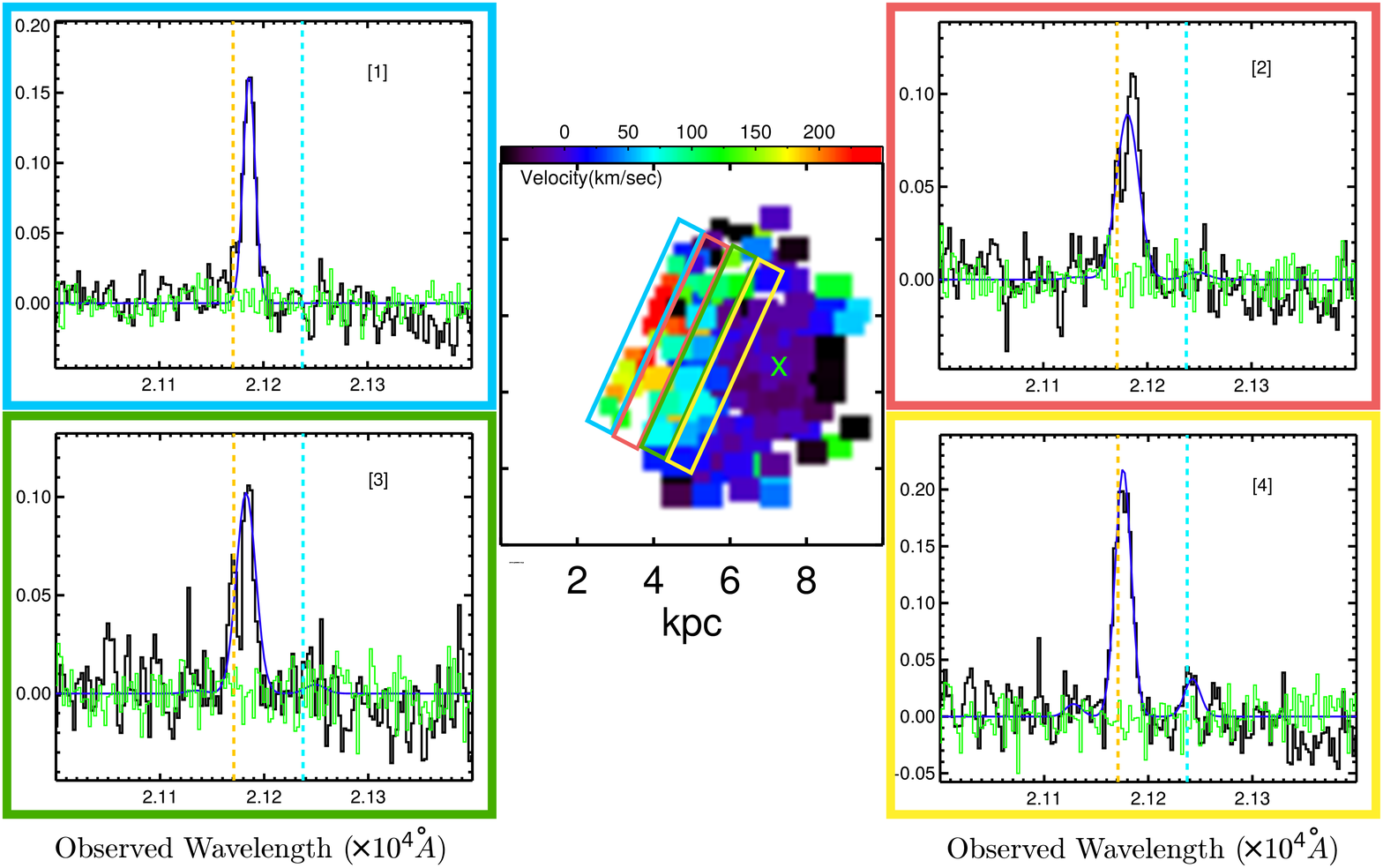}
\caption{Derived source-plane 2D rotational velocity map and integrated 1D spectra of different source-plane regions using our forward-adaptive co-adding technique. The kinematical center ($z = 2.225$), shown by a green cross on 2D velocity map, is chosen to lie at the peak of H$\alpha$ surface brightness distribution. The integrated spectra of different spaxels show a clear double peak as we move away from the center. The raw spectrum is plotted in black; RMS of sky residuals are plotted in green and gaussian fits to H$\alpha$ and [N\,{\sc ii}\rm] are shown in blue. Yellow and Cyan vertical lines in Panels 1-4 indicate the location of redshifted H$\alpha$ and red [N\,{\sc ii}\rm] emission lines at $z = 2.225$ respectively.}
 \label{fig:fig6}
\end{figure*}

We notice that a mask size of upto 5$\times$5 pixels is sufficient to obtain an emission line fitting result of H$\alpha$ SNR $\ge$ 5$\sigma$ in different regions of the source. Using our adaptive forward modeling technique, we exploit the lensing amplification to optimize the spatial resolution in the source plane. This allows us to confidently analyse the dynamics of the lensed target which was previously limited only to the high SNR regions of the image plane. 

In Figure~\ref{fig:fig5}, we compare the source reconstructions from our forward modeling technique with the traditional ray tracing methodology. We ray-trace the image plane fitted 2D H$\alpha$ intensity map for multiple image a2 (as shown in Figure~\ref{fig:fig2}) and H$\alpha$ SNR map back to the source plane and compare them to our coadded source plane reconstruction. As apparent in Figure~\ref{fig:fig5}, we are able to obtain a SNR improvement by a factor $>$ 5 in the source plane. The improved SNR is achievable due to adaptively binning the image plane pixels, where the number of binned spaxels varies on the location of their corresponding source position. The forward modeling enables a relatively uniform PSF throughout the source-plane as compared to the variable PSF attained through traditional ray-tracing. We note that the image plane RMS of $0.\!\!^{\prime\prime}06$ due to the lens model will translate into only small offsets between the co-added source plane locations of a1 and a2 which will not affect the reconstruction significantly.

\section{Results}
In the sections below, we discuss the source plane morphological and kinematical properties of cswa128.
\subsection{Source Plane Morphology}
We reconstruct the continuum image from NIRC2 observations using \textit{cleanlens} task in {\lenstool} and present the source-plane morphology in Figure~\ref{fig4:sub1}. The reconstructed continuum intensity distribution, dominated by the stars in the galaxy, shows a clumpy structure. On the other hand, 2D source-plane H$\alpha$ map (Figure~\ref{fig:fig5}) generated as a result of our forward modelling technique also shows a clumpy but more extended emission compared to the reconstructed continuum image of the lensed galaxy. As emphasized by previous studies \citep[e.g][]{Elmegreen09, Forster11, Girard18}, clumpy morphology is commonly observed for high-$z$ galaxies irrespective of their kinematical properties. 

Our improved source-plane H$\alpha$ morphology allows us to investigate the spatial distribution of the ionized gas emission in greater detail than before. We find that the reconstructed morphology of the ionized gas emission as traced with H$\alpha$ differs significantly from the continuum emission, tracing the stellar structure of the lensed galaxy. As shown in Figure~\ref{fig:fig5}, we observe a clear mismatch in the location of the surface brightness peaks in the continuum and H$\alpha$ maps. The spatial offsets between stellar continuum and ISM H$\alpha$ gas have been previously observed in local luminous and ultra-luminous infrared galaxies dominated by merging events  \citep[e.g.][]{Colina05, Rodrig11, Barrera15}.
\subsection{Kinematics}
The mismatch between stellar and ionised gas emission in cswa128 provides promising evidence for an ongoing merger. However, it is difficult to draw any conclusions about the disk/merger nature of high-$z$ galaxies through morphology alone. Combining kinematics with morphology offers a distinct advantage to distinguish merger from isolated disks \citep[e.g.][]{Shapiro08, Rod17, Yuan17}. 

In this work, we acquire a much deeper understanding of the kinematical properties of cswa128 with the improved SNR of H$\alpha$ and [N\,{\sc ii}\rm] emission lines in the source plane, thanks to our forward modeling approach of combining the data from two different images of the galaxy. Our analysis indicates that the system is most likely undergoing a merging process. This result is consistent with the previous work by \citetalias{Leethochawalit16}. This section outlines the different observations that led us to the above conclusion. 

Figure~\ref{fig:fig6} shows the 2D velocity map obtained as a result of our co-adding technique described in Section~\ref{sec:forward_approach}. The velocity gradient is highly inconsistent with a well-ordered rotating system. During the manual inspection of spectra, we notice a clear hint of double-peaked H$\alpha$ emission in many spaxels, especially at the edges of the galaxy, indicating the presence of a merging pair in the system. This double peak signature was invisible in the previous study of this lensed system. We extract a 1D spectrum of different regions of the source-plane to show the presence of double-peaked H$\alpha$ emission. As shown in Figure~\ref{fig:fig6}, the double-peaked emission is clearly visible as we move away from the kinematical center of the galaxy which is simply chosen to lie at the peak of H$\alpha$ brightness distribution.

To further investigate the nature of the double peak, we refit H$\alpha$ and [N\,{\sc ii}\rm] emission lines, in the spectra of previously created masks but this time using double-gaussian profile each representing a different component of the emission. We refer to the two components as blue and red, in H$\alpha$ and [N\,{\sc ii}\rm] lines associated with every spaxel in the source plane. Because certain regions of the source-plane show quite a strong emission in one component than the other (see  panels 1,4 in Figure~\ref{fig:fig6}), we define redshift ranges for the two components to allow the fitting routine to identify the correct one in the spectrum. Assisted by the initial 2D velocity map in Figure~\ref{fig:fig6}, we first associate the blue component with the disc and allow it to vary within $\pm 80$ km/sec of the kinematical center ($z \sim 2.225$). However on detailed investigation of individual spaxels at the edges of the galaxy, we find some spaxels showing emission bluer than the expected range, so we further extend the lower limit to -150 km/sec. Because the blue component gets weaker as we move away from the center of the galaxy, our preliminary analysis using a single component fit will only be dominated by the much stronger redder emission. The red component is used to fit any significant emission other than the blue component; quantitatively extending as $\pm 150$ km/sec of the redshift $z = 2.228$. Like before, [N\,{\sc ii}\rm] (blue/red) SNR cutoff is set to 3$\sigma$. After visually inspecting the edgy spaxels, we identify H$\alpha$-blue SNR $\ge$ 3$\sigma$ and H$\alpha$-red SNR $\ge$ 5$\sigma$ as genuine detections. 

\begin{figure*}
\centering
\includegraphics[scale=0.45,angle=0]{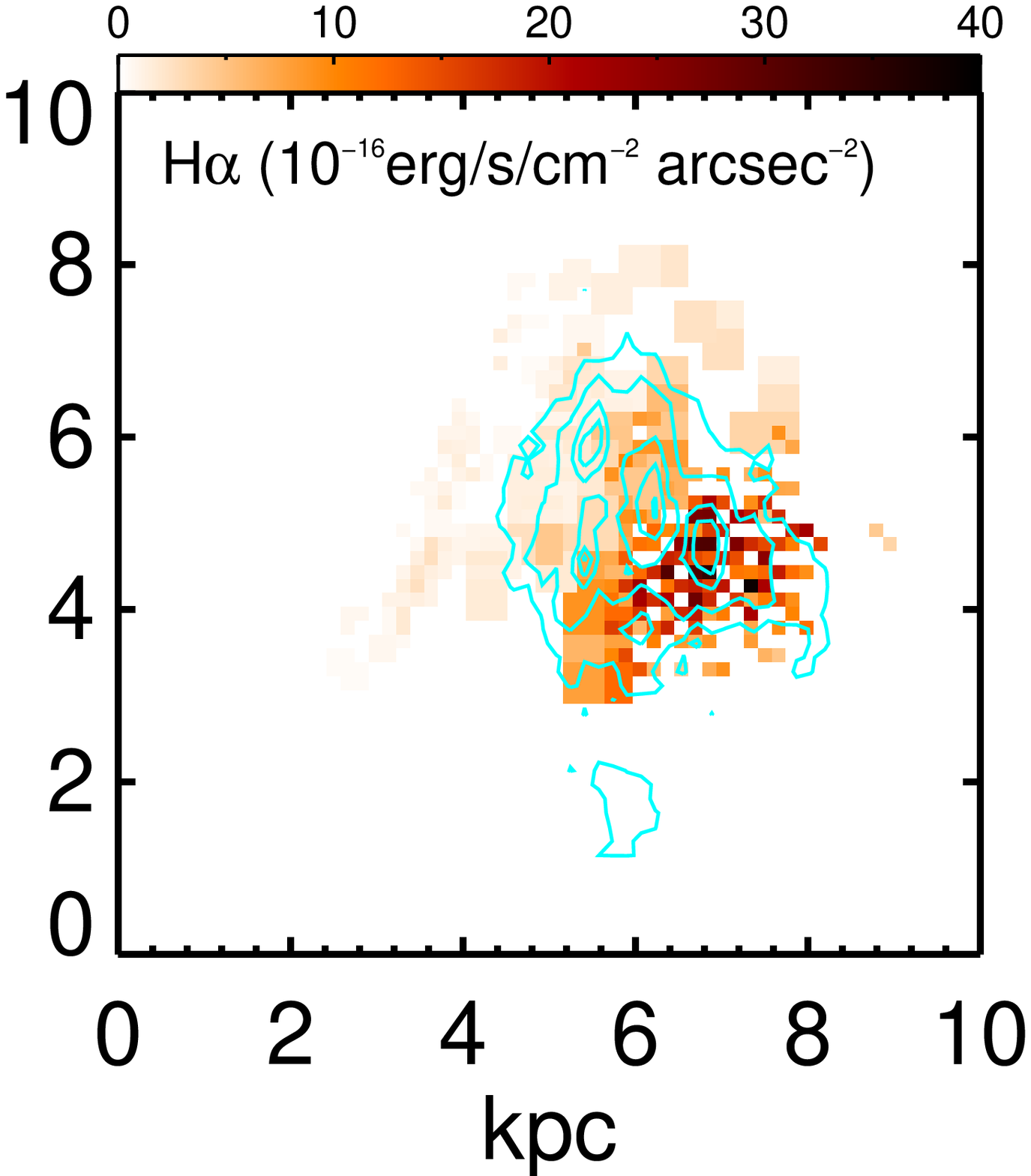}
\includegraphics[scale=0.45,angle=0]{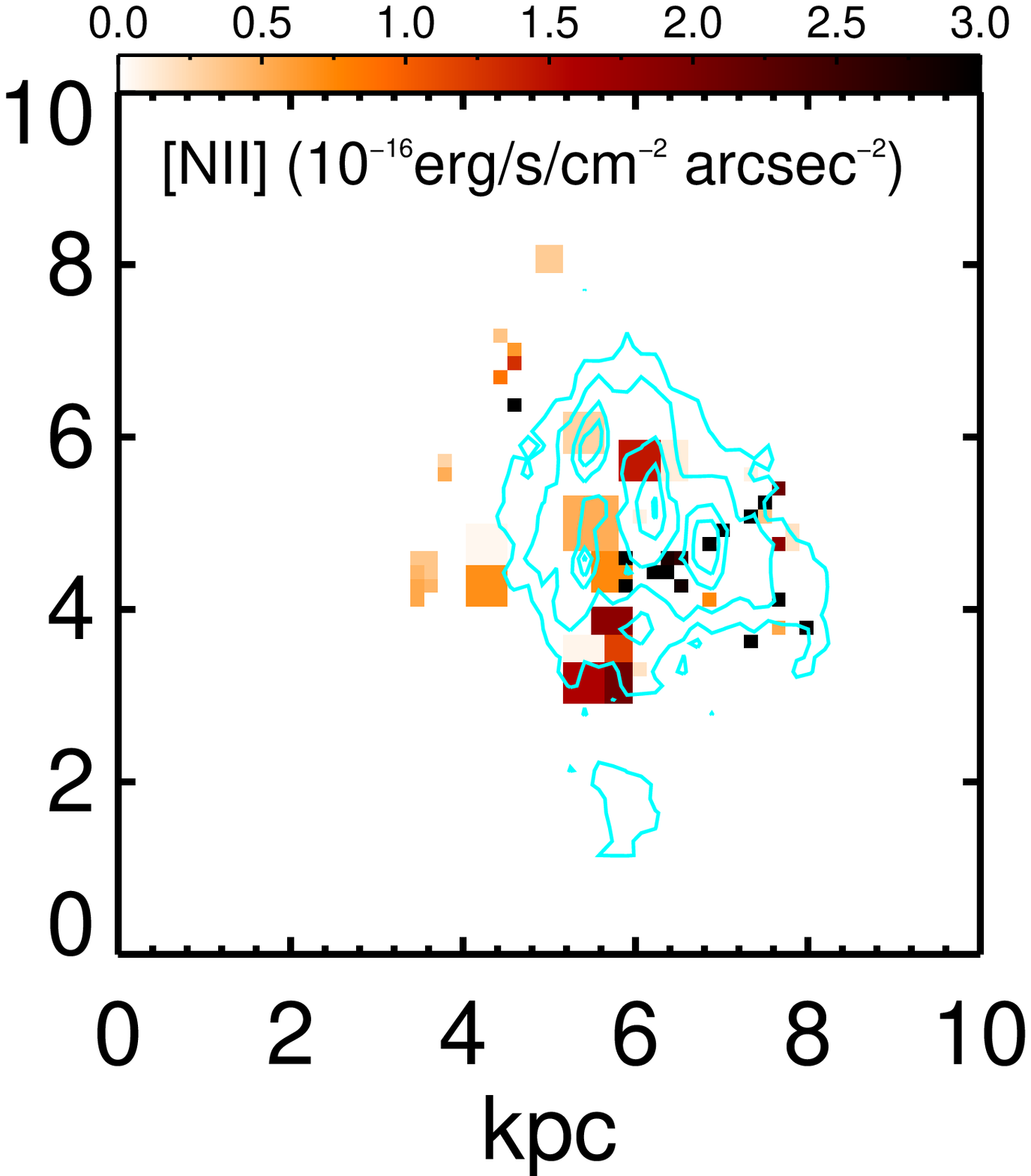}
\includegraphics[scale=0.45,angle=0]{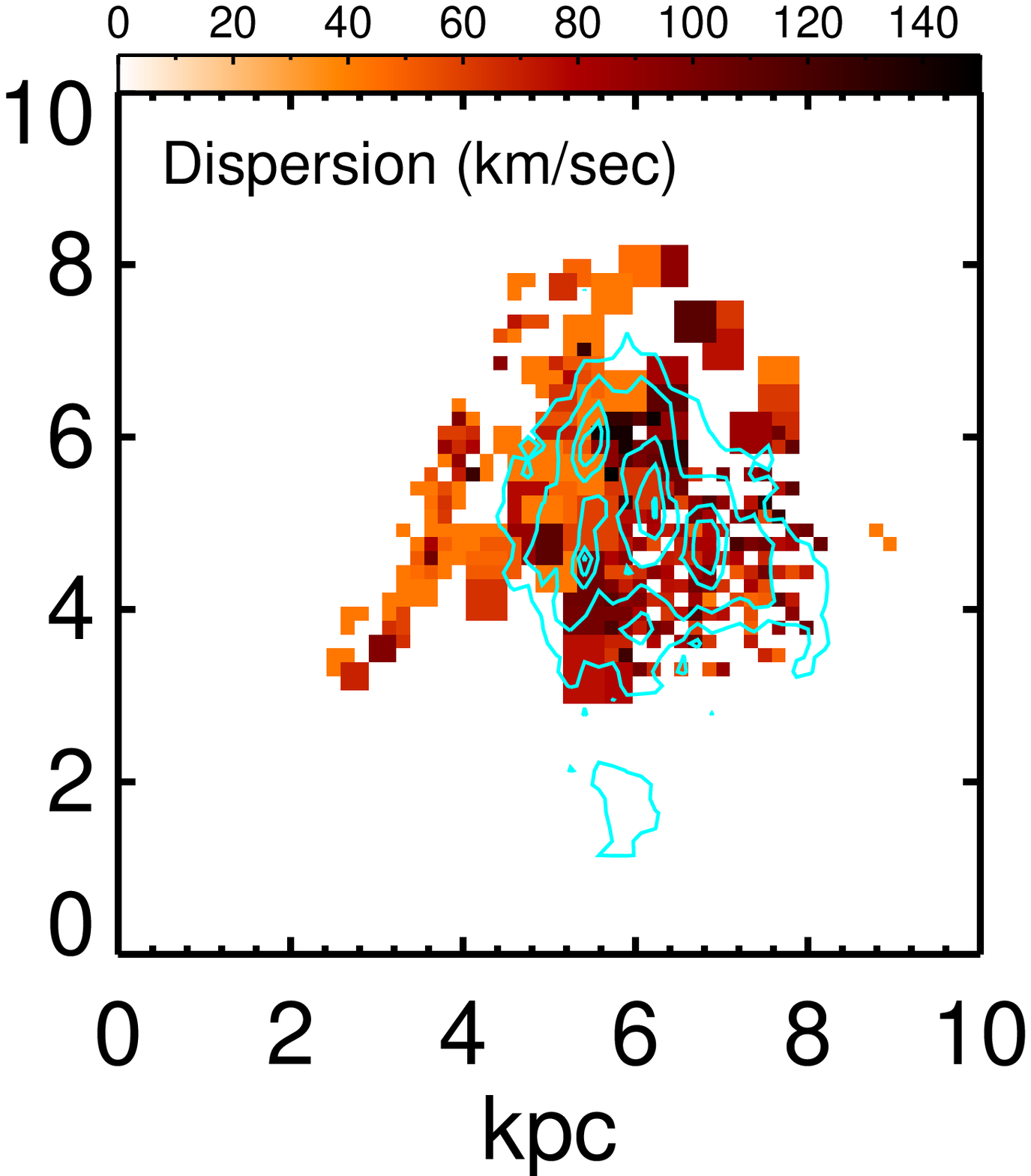}
\includegraphics[scale=0.45,angle=0]{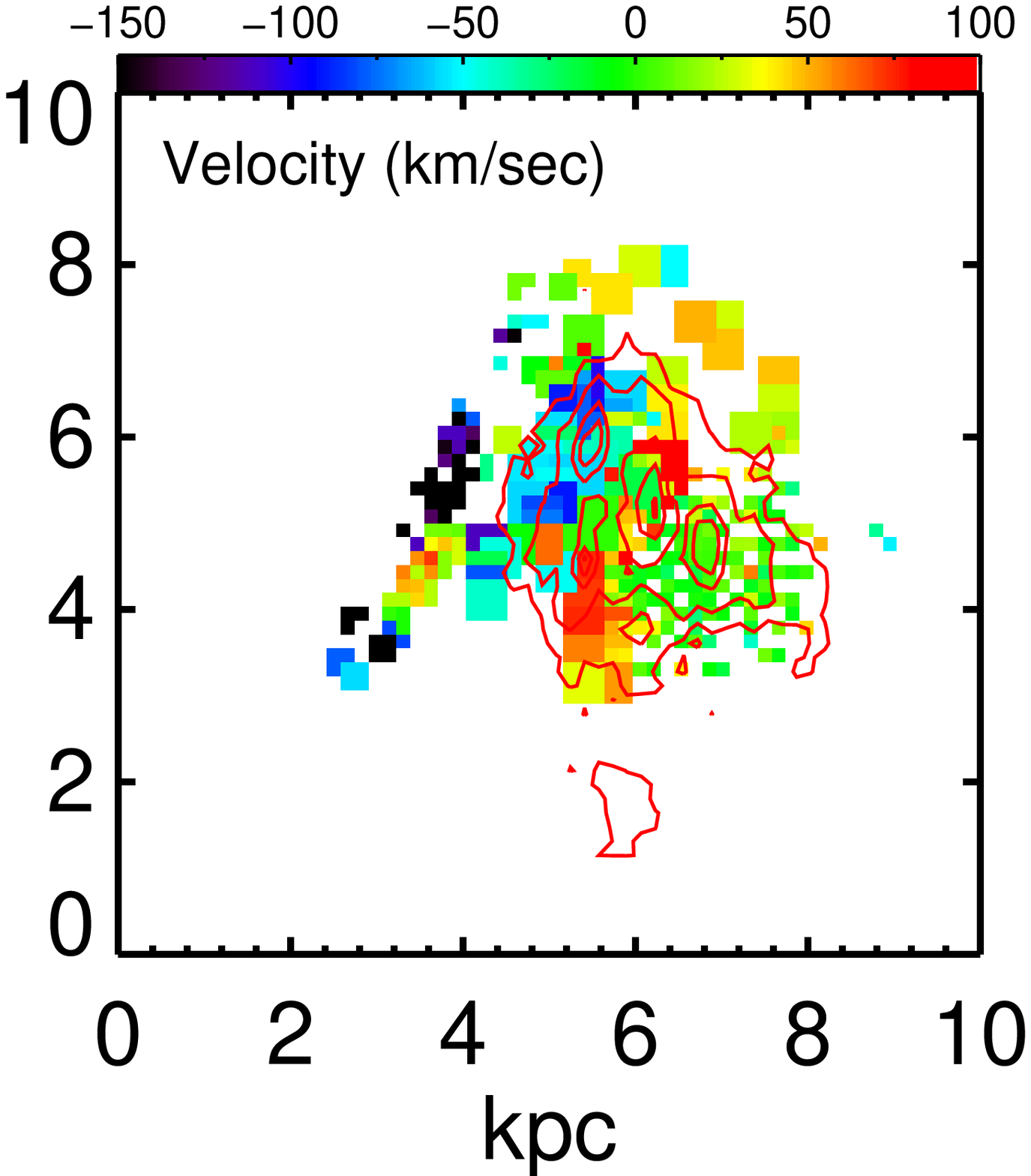}
\caption{Derived source plane 2D maps for blue component of emission. The rotational velocity is calculated at the rest-frame of $z = 2.225$ and with respect to the kinematic center (green cross in Figure~\ref{fig:fig6}). Contours are the same as Figure~\ref{fig:fig4}\label{fig:fig7}.}
\end{figure*}

\begin{figure*}
\centering
\includegraphics[scale=0.45,angle=0]{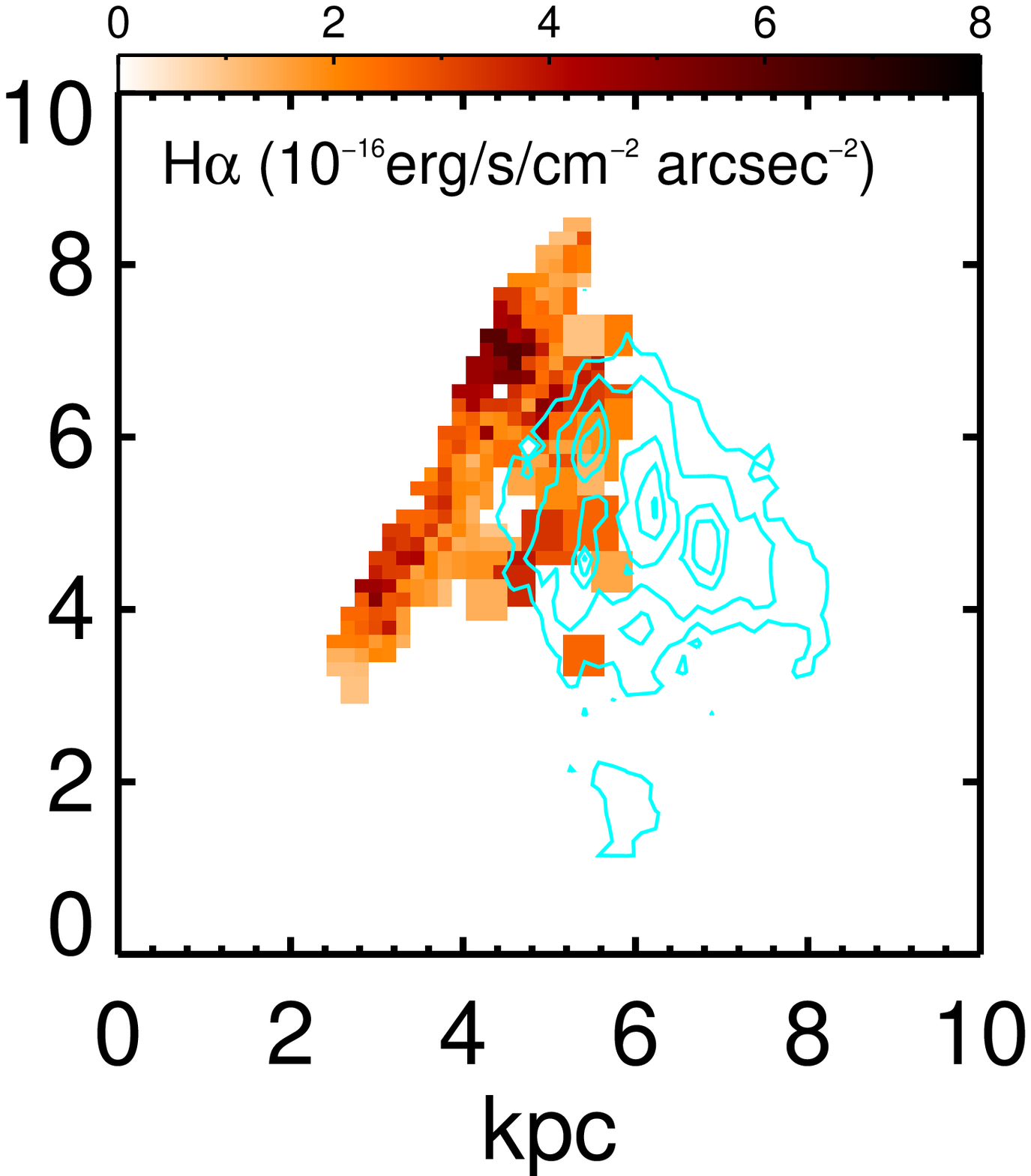}
\includegraphics[scale=0.45,angle=0]{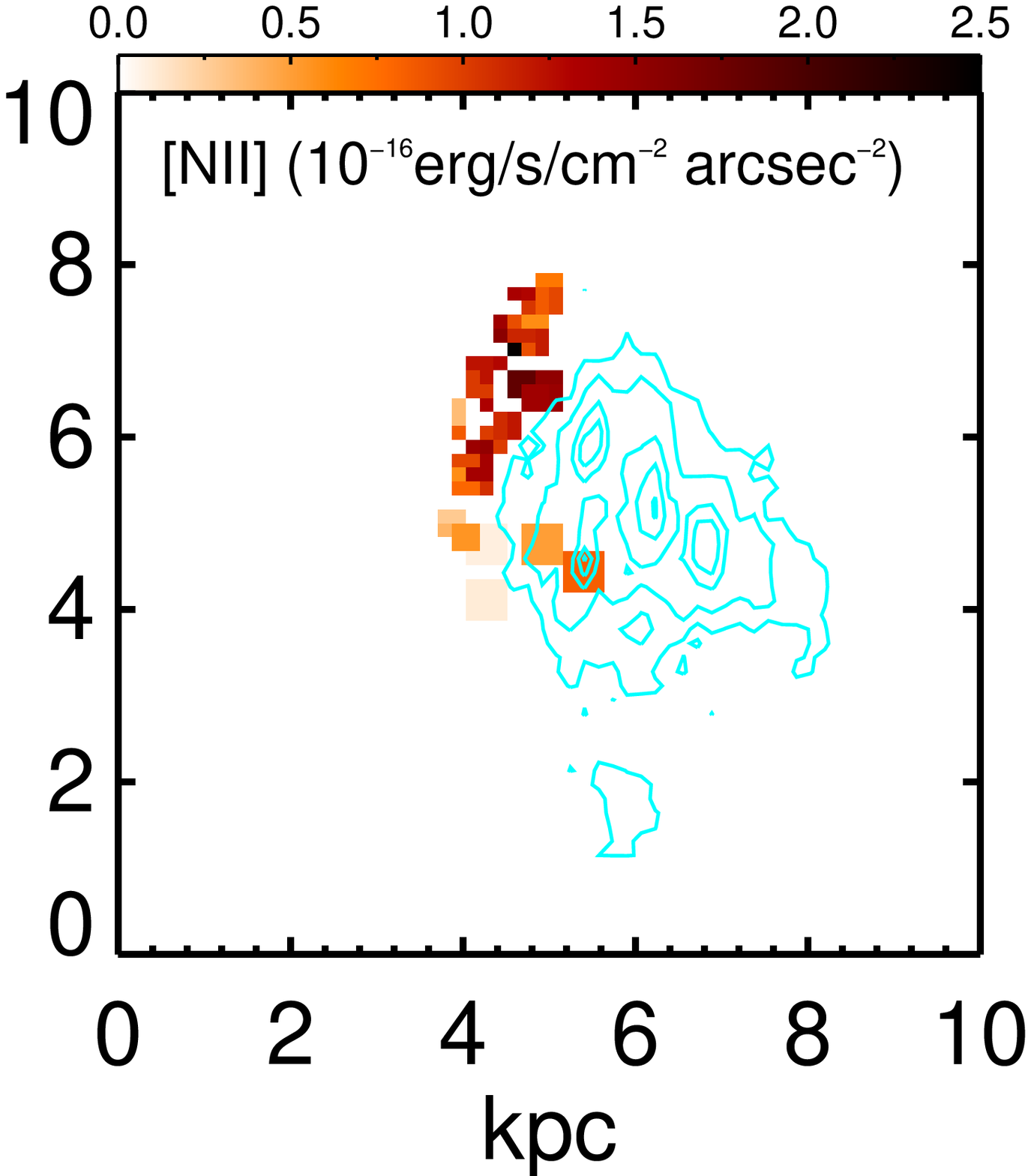}
\includegraphics[scale=0.45,angle=0]{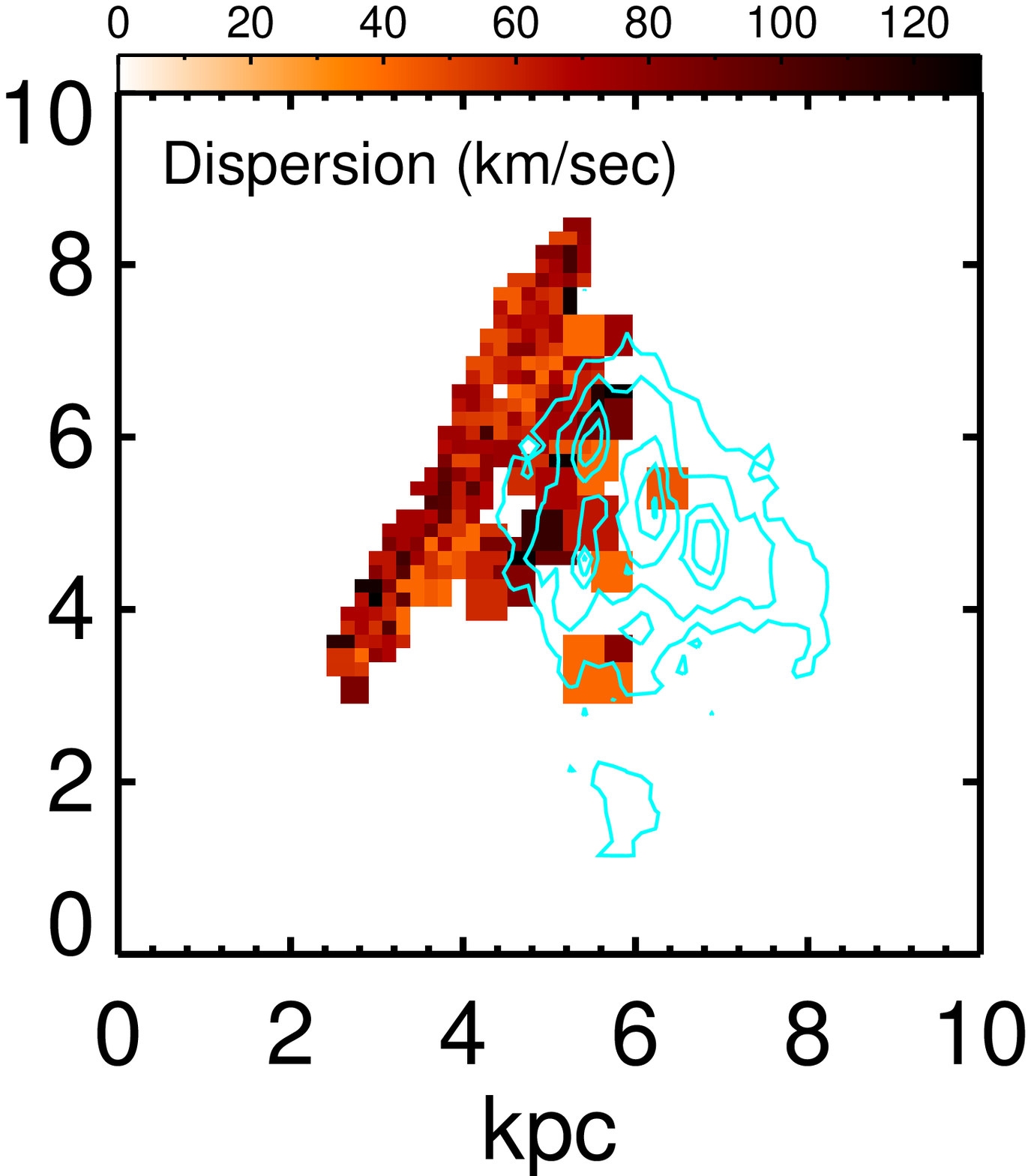}
\includegraphics[scale=0.45,angle=0]{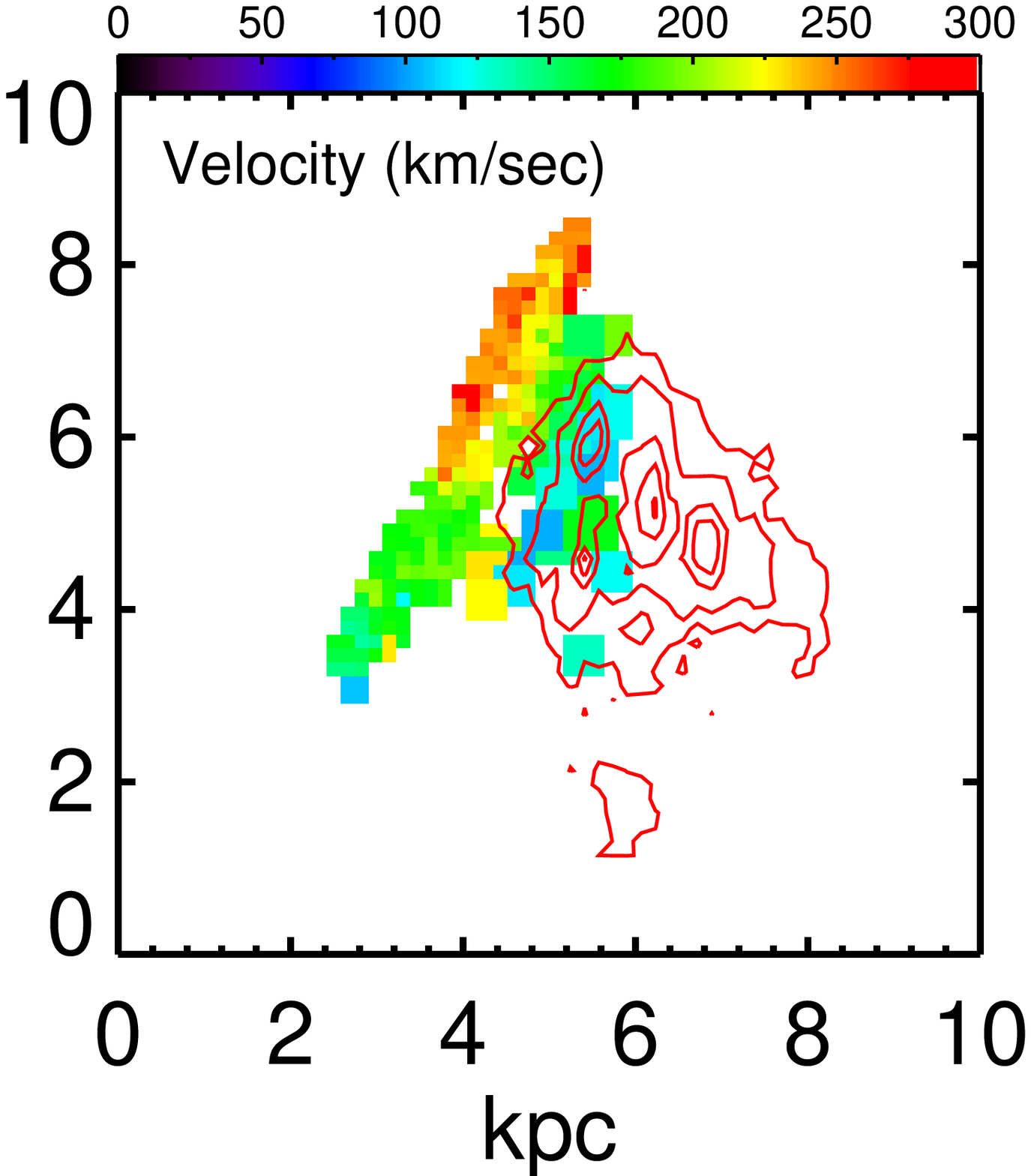}
\caption{Derived source plane 2D maps for red component of emission. The rotational velocity is calculated at the rest-frame of $z = 2.225$ and with respect to the kinematic center (green cross in Figure~\ref{fig:fig6}). Contours are the same as Figure~\ref{fig:fig4}\label{fig:fig8}.}
\end{figure*}

The derived 2D maps for blue and red component are shown in Figure~\ref{fig:fig7} and Figure~\ref{fig:fig8} respectively. We find that the source-plane 2D velocity field for the blue component is highly disturbed and remains inconsistent with a rotating disk. The average velocity dispersions for the blue and red component are 69.9 $\pm$ 25.0 km/sec and 70.2 $\pm$ 21.5 km/sec respectively. This is in agreement with the observations of high-redshift mergers, further strengthening the merger scenario in cswa128. 

\subsection{Rest-Frame Emission Line Ratios}
\begin{figure*}
\centering
\includegraphics[scale=0.45,angle=0]{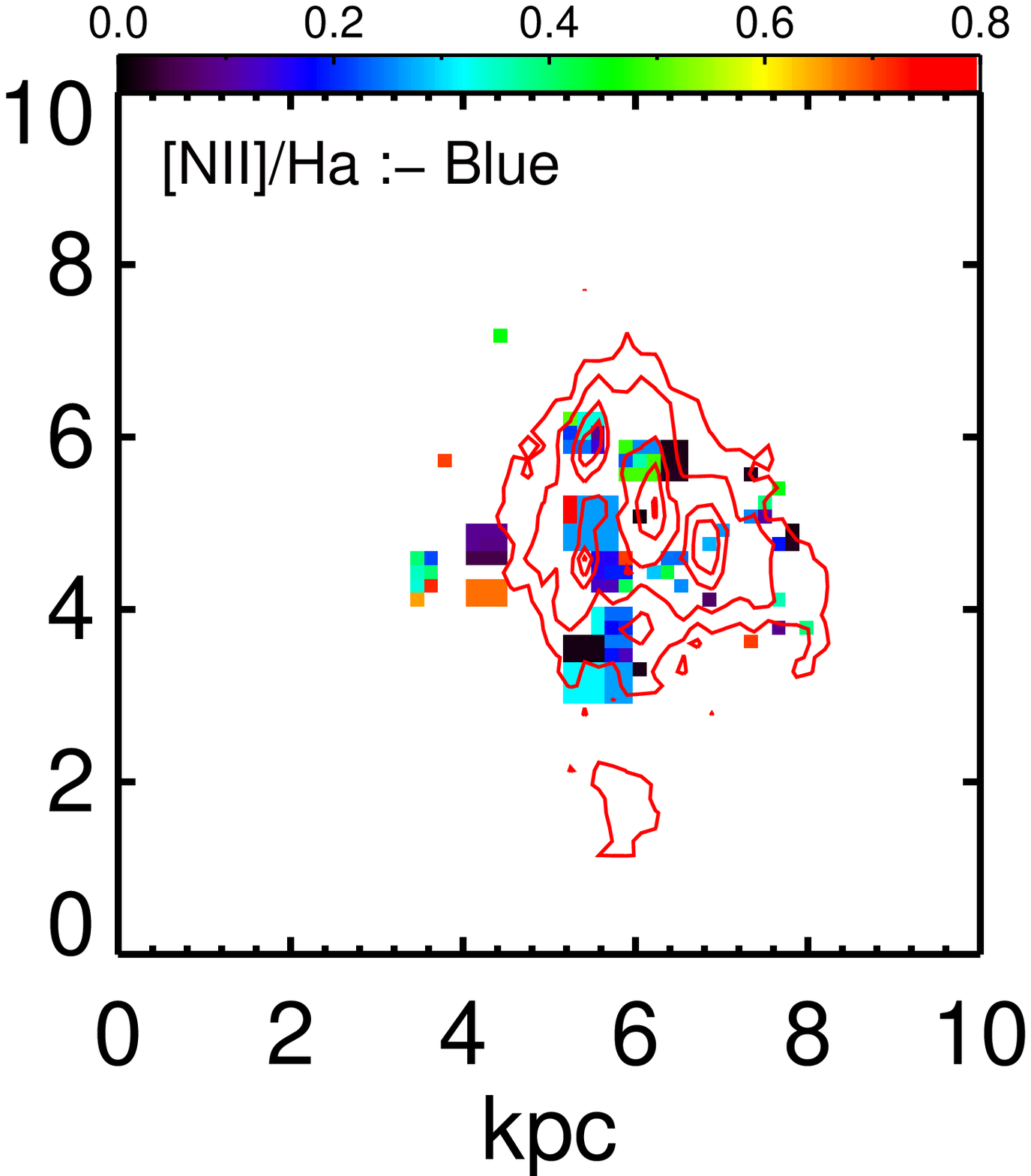}
\includegraphics[scale=0.45,angle=0]{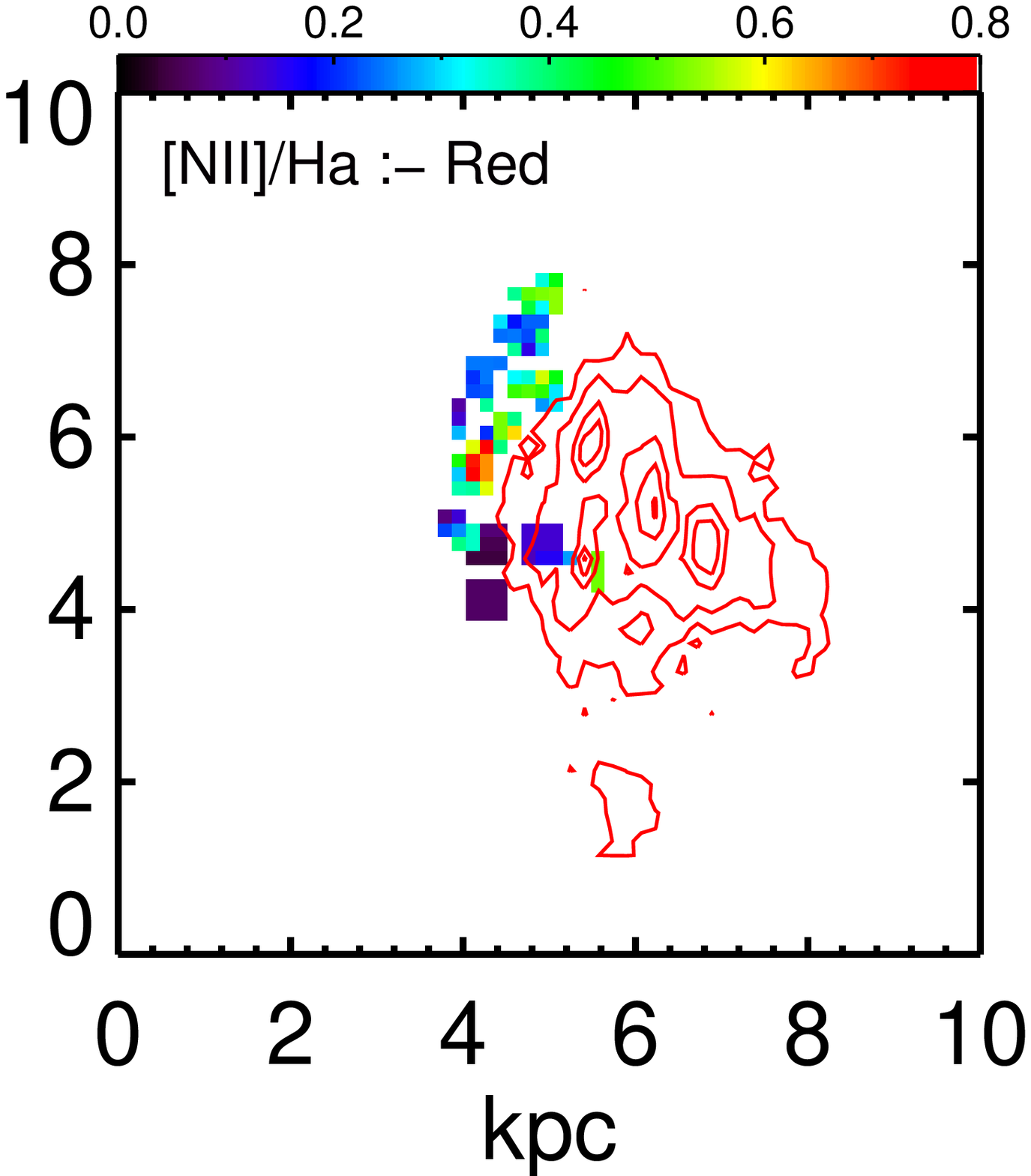}
\includegraphics[scale=0.45,angle=0]{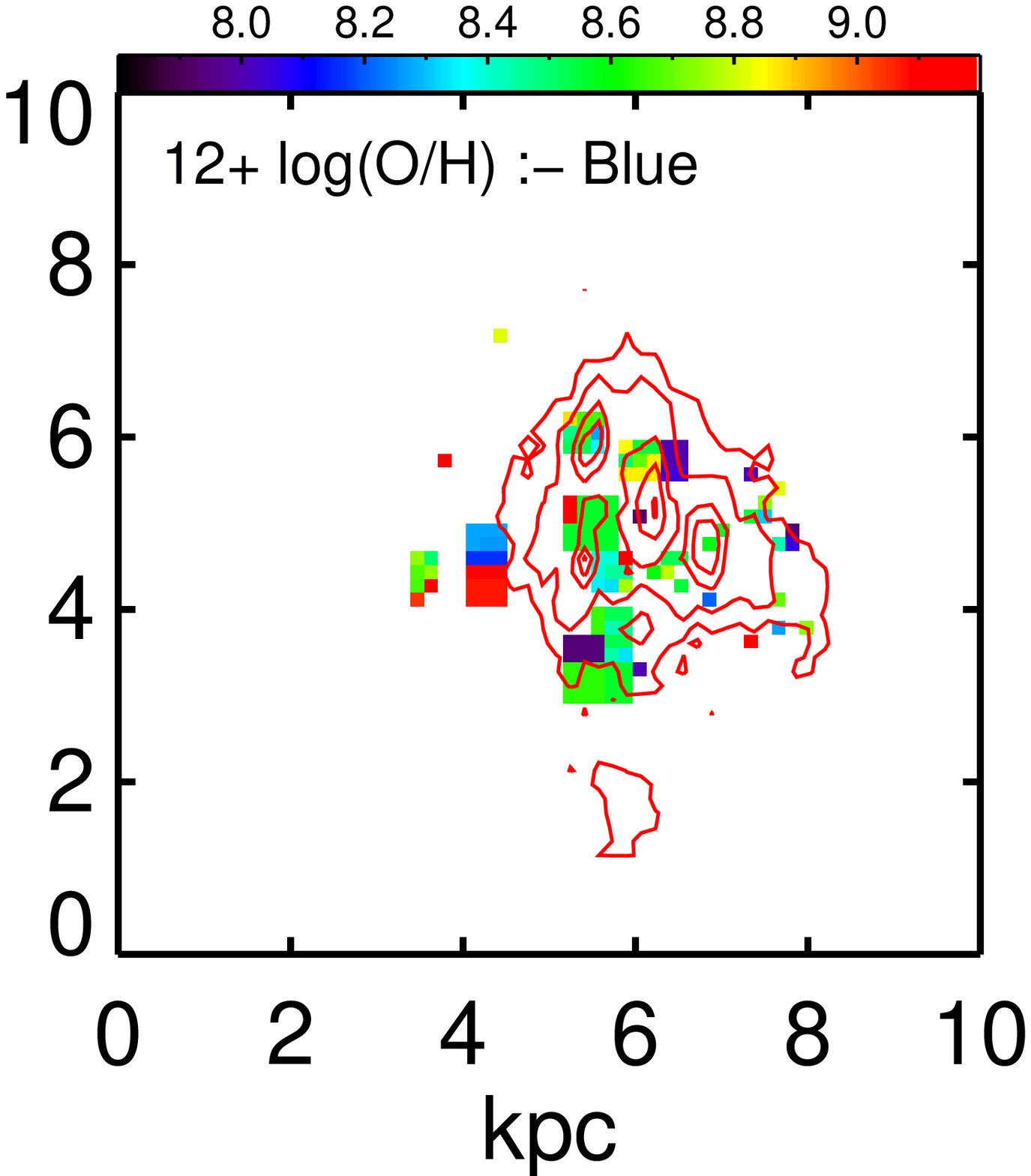}
\includegraphics[scale=0.45,angle=0]{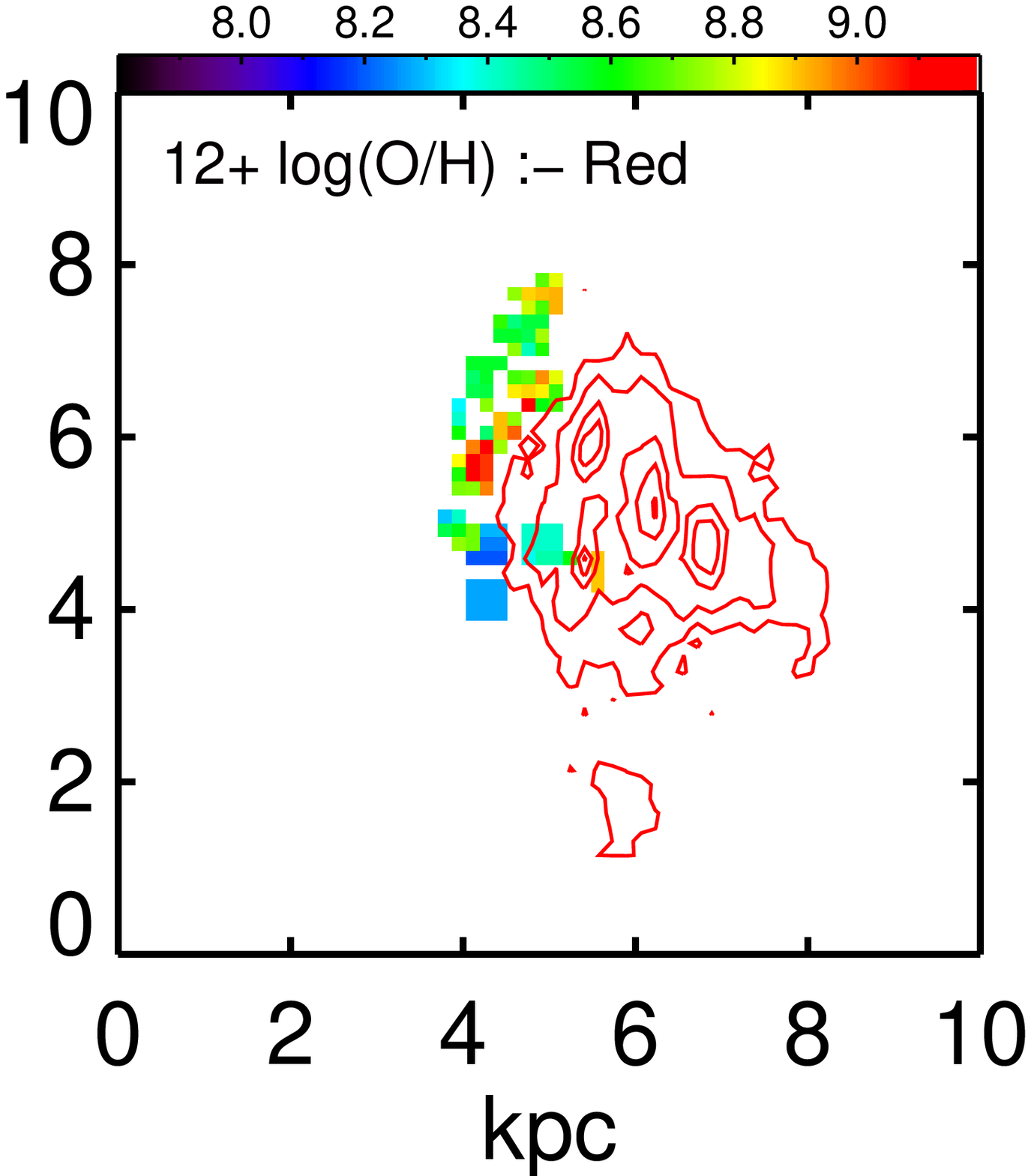}
\caption{Derived 2D source-plane [N\,{\sc ii}\rm]/H$\alpha$ and metallicity maps for blue and red components of emission. Contours are the same as Figure~\ref{fig:fig4}\label{fig:fig9}.}
\end{figure*}

Figure~\ref{fig:fig9} shows the derived 2D source-plane [N\,{\sc ii}\rm]/H$\alpha$ maps for the blue and red component of emission. The average log([N\,{\sc ii}\rm]/H$\alpha$) for the blue and red components are respectively -0.56 $\pm$ 0.31 and -0.51 $\pm$ 0.29. With these ratios alone, we can not rule out contaminations from shocks or AGN \citep{Kewley13}. We therefore advise readers to interpret metallicity of this system with caution. Further observations including more line ratios such as [O\,{\sc iii}\rm]/H$\beta$ and [S\,{\sc ii}\rm] will help to provide a more robust diagnostic for metallicity.

Further, we use the calibration of \cite{Pettini04} to convert [N\,{\sc ii}\rm]/H$\alpha$ into metallicity. The correlation between metallicity and N2 ($\equiv$ log([N\,{\sc ii}\rm]$\lambda6583$/H$\alpha$)) used is as follows : 
\begin{equation}
\mathrm{12 + log(O/H) = 8.90 + 0.57 \times N2 }   
\end{equation}
As shown in Figure~\ref{fig:fig9} we observe flat metallicity gradients, a signature commonly observed in merging galaxies \citep{Kewley10, Wuyts16, Cortijo17, Wang17}. We notice that our metallicity distribution maps are limited by the low SNR of [N\,{\sc ii}\rm] emission lines. Deeper IFS data is needed to better resolve [N\,{\sc ii}\rm] and perform a detailed analysis of emission line ratios. 

\subsection{Spectral Energy Distribution fitting, Stellar Mass and Star formation Rate}
We use the software {\lephare} \citep{Ilbert10} to perform SED fitting of the lensed system in cswa128. {\lephare} computes the best-fit SED based on population synthesis models by \cite{Bruzual03}. It uses a $\chi2$ minimization procedure to provide the best-fit parameters such as stellar mass, SFR and extinction value. We fix the redshift to 2.22 and use the attenuation law by \cite{Calzetti00} and IMF by \cite{Chabrier03}. We use E (B -V) ranging from 0 to 2 and an exponentially decreasing SFR ($\propto$ e$^{-t/ \tau}$) with $\tau$ varying between 0 and 13 Gyr. We estimate the best fit stellar mass  as 10$^{10.4\pm 0.3}$ M$_{\odot}$ and extinction E(B-V) = 0.4. The total SFR from the best-fit SED is 361 $\pm$ 411 M$_{\odot}$ yr$^{-1}$. All values have been corrected for lensing flux magnification. Because of limited photometric bands, {\lephare} obtains high uncertainties in SFR estimates from the SED fitting. Therefore we calculate dust-uncorrected SFR using total H$\alpha$ emission as 201.92 $\pm$ 20.78 M$_{\odot}$ yr$^{-1}$. The total dust uncorrected SFR of cswa128 is $\sim$ 3.5 times higher than typical star forming galaxies lying on the $z \sim 2$ mass-SFR relation \citep[e.g.][]{Zahid12} of the main sequence. This is in agreement with the merger scenario also supported by the turbulent kinematics, double-peaked emission lines and the offsets between ionised gas and stellar continuum.

\section{Conclusion and Future Work}\label{sec:discussion}
Gravitationally lensed systems offer us the unique opportunity to study galactic dynamics at an unprecedented physical resolution in the source plane. Such a fine spatial resolution will not be achievable even with future instruments such as the NIRSpec aboard JWST or GMTIFS on GMT.\footnote{The spatial sampling of NIRSpec is 100mas, and AO aided GMTIFS will have a spatial resolution of 10-25 mas.\label{footnote:gmt}} Ongoing lensing surveys will increase the sample of lensed galaxies by orders of magnitude \citep[e.g. LSST,][]{Oguri10,Marshall10}. To prepare ourselves for interpreting large samples of lensed galaxies, it is of crucial importance to continue studying individual cases like cswa128 in great detail in order to better understand the systematics of lens models and capture the important physical properties of  high-$z$ galaxies that are otherwise unattainable without lensing.

For example, the fraction of mergers in star-forming galaxies at $z \sim 1-3$ has long been debated. In this redshift regime, merger fractions show a strong dependence with the classification scheme used  to classify merger signatures in high-$z$ observational data \citep[e.g.][]{Hung15, Molina17}. Predictions from high resolution simulations show that current observations significantly underestimate the occurrence of mergers in typical star forming galaxies  at $z \ge$1 \citep{Rodrigue15}. Gravitational lensing coupled with high resolution imaging and spectrographic observations have the potential to unambiguously classify a galaxy as a merger or disk dominated as demonstrated in this work. In fact, without gravitational lensing,  a galaxy like cswa128 would easily be misclassified as a typical clumpy SF galaxy at $z \sim$ 2.

 The reliability of source-plane properties of lensed galaxies depends strongly on the accuracy of the lensing mass models used to interpret the high resolution observations. In this work, we use high resolution imaging data to identify additional substructures and therefore construct a more robust map of the lensing potential of the galaxy group in cswa128 at $z = 0.214$. 

However, in order to fully exploit the potential of these natural gravitational telescopes, it is important to assimilate information from multiple images of the lensed target. Along with a high resolution lens model, we present an adaptive forward modelling approach to combine observational data sets on different instances of the lensed arc. We resolve the structure and kinematics of the lensed target at scales of $\sim$170 pc in the source plane. We are able to detect different velocity components of [N\,{\sc ii}\rm] and H$\alpha$ emission lines using this approach that are not seen in traditional source-plane coadding methods. 
We propose merger as the most plausible scenario in cswa128 on the basis of following observations:
\begin{enumerate}
\item {\bf Morphology:} Offset between the clumpy stellar emission compared to the rather extended H$\alpha$ morphology.
\item {\bf Kinematics:} Chaotic velocity distribution with a clear double-peaked H$\alpha$ emission possibly representing different components of merger. 
\item {\bf Chemical gradient:} Flat metallicity gradients derived from [N\,{\sc ii}\rm]/H$\alpha$ emission line ratios.
\end{enumerate}
The results are consistent with the previous analysis by \cite{Leethochawalit16}. However, with the improved SNR of the emission lines in the source plane, we find more conclusive evidence of an ongoing merger in cswa128.

The forward modeling technique presented in this paper has been developed specifically for this lensed target and required considerable manual intervention, especially in the construction of source masks. However, being computationally inexpensive, this approach offers a very quick and effective way of combining the data from multiple images to enhance the SNR of emission line maps in the source plane. In our follow-up work, we extend this forward source modeling approach to a fully automated inversion of extended lensed images in a Bayesian manner. We present an algorithm in {\lenstool} to obtain a discretized surface brightness distribution in the source plane for a fixed mass profile of the lens. We use a pixelized grid and Bayesian MCMC optimization algorithm implemented in {\lenstool} to employ a forward modeling approach to deconvolve PSF effects from the source profile. The algorithm allows higher flexibility than traditional parametric ways of modelling source galaxies thus allowing for unbiased reconstructions of irregular source morphologies. Moreover, this method will be widely applicable in studying a variety of lensed systems that will become available with the future instruments and lensed surveys \citep{Rydberg18, Agnello18}.

We thank the comments of the anonymous referee, which helped to improve the paper. This work is based on data obtained at the W. M. Keck Observatory. We are grateful to the Keck Observatory staff for assistance with our observations, especially Randy Campbell and Jason. This research was supported by the Australian Research Council Centre of Excellence for All Sky Astrophysics in 3 Dimensions (ASTRO 3D), through project number CE170100013. We thank Richard Ellis for sharing his OSIRIS data. SS thanks useful discussions with Prasun Dutta, GEARS3D and CALENDS group. SS is also thankful to Surya Narayan Sahoo for technical help in formatting some of the figures in the paper. The authors wish to recognize and acknowledge the very significant cultural role and reverence that the summit of Mauna Kea has always had within the indigenous Hawaiian community.

\bibliographystyle{mnras}
\bibliography{bib_soniya}

\end{document}